\documentclass[]{aa}  
\usepackage{graphicx}
\usepackage{txfonts}
\usepackage{subfigure}
\usepackage{natbib}
\usepackage{longtable}
\usepackage{comment}
\usepackage{hyperref}

\usepackage{ulem}

\usepackage{color}
\usepackage{amstext}

\begin{document}

\newcommand{\kms}{km~s$^{-1}$}
\newcommand{\ms}{m~s$^{-1}$}
\newcommand{\cms}{cm~s$^{-1}$}
\newcommand{\halp}{H$\alpha$}
\newcommand{\msun}{$\rm M_{\odot}$}
\newcommand{\etal}{et~al.~}
\newcommand{\vsini}{$v~sin~i$}
\newcommand{\ctemp}{$^{\circ}$C}
\newcommand{\ktemp}{$^{\circ}$K}
\newcommand{\be}{\begin{equation}}
\newcommand{\ee}{\end{equation}}
\newcommand{\bd}{\begin{displaymath}}
\newcommand{\ed}{\end{displaymath}}
\newcommand{\bi}{\begin{itemize}}
\newcommand{\ei}{\end{itemize}}
\newcommand{\bfig}{\begin{figure}}
\newcommand{\efig}{\end{figure}}
\newcommand{\bc}{\begin{center}}
\newcommand{\ec}{\end{center}}
\newcommand{\hii}{{H\scriptsize{II}}}
\newcommand{\vlsr}{v$_{\mathrm{LSR}}$}
\newcommand{\teff}{T$_{\mathrm{eff}}$}
\newcommand{\vtur}{V$_{\textrm{\tiny{tur}}}$}
\newcommand{\vexp}{V$_{\textrm{\tiny{exp}}}$}
\newcommand{\vinfall}{V$_{\textrm{\tiny{infall}}}$}
\newcommand{\coa}{$^{12}\mathrm{CO}$}
\newcommand{\cob}{$^{13}\mathrm{CO}$}
\newcommand{\coc}{$\mathrm{C}^{18}\mathrm{O}$}
\newcommand{\sioa}{$^{28}\mathrm{SiO}$}
\newcommand{\siob}{$^{29}\mathrm{SiO}$}
\newcommand{\sioc}{$^{30}\mathrm{SiO}$}
\newcommand{\lsun}{L$_{\odot}$~}
\newcommand{\lfir}{L$_{\textrm{\tiny{FIR}}}$}
\newcommand{\agua}{$X_{\textrm{\tiny{H$_2$O}}}$}
\newcommand{\ratioop}{o$/$p}
\newcommand{\ratiosept}{$X_{\textrm{\tiny{$^{18}$O$/$$^{17}$O}}}$}
\newcommand{\ratiohuit}{$X_{\textrm{\tiny{$^{16}$O$/$$^{18}$O}}}$}
\newcommand{\lsol}{L$_\odot$\,}
\newcommand{\Msol}{M$_\odot$\,}
\newcommand\omicron{o}

\newcommand{\mm}[1]{\textcolor{green}{MM: #1}}  
\newcommand{\lm}[1]{\textcolor{blue}{#1}}
\newcommand{\hw}[1]{\textcolor{magenta}{#1}}

\def\etal{et al.$\;$}
 

\def\kms{km\thinspace s$^{-1}$}
\def\Lsun{L$_\odot$}
\def\Msun{M$_\odot$}
\def\ms{m\thinspace s$^{-1}$}
\def\percc{cm$^{-3}$}
\title{SiO maser polarization and magnetic field in evolved cool stars}

\titlerunning{SiO maser Polarization and magnetic field in evolved cool stars}

\subtitle{}

\author{L. Marinho,
        \inst{1}
        F. Herpin,
        \inst{1}   
        H. Wiesemeyer,
        \inst{2}
        \and
        A. L\'opez Ariste
        \inst{3}     
        \and
        A. Baudry
        \inst{1}
        \and
        A. Asensio Ramos
        \inst{4}
        \and
        A. Lèbre
        \inst{5}
        \and
        P. Mathias
        \inst{6}
        \and
        M. Montargès
        \inst{7}
}

\institute{
Laboratoire d'astrophysique de Bordeaux, Univ. Bordeaux, CNRS, B18N, all\'ee Geoffroy Saint-Hilaire, 33615 Pessac, France.
        \email{louise.marinho@u-bordeaux.fr}
\and
Max-Planck-Institut f\"ur Radioastronomie, Auf dem H\"ugel 69, 53121 Bonn, Germany
\and
IRAP, Universit\'e de Toulouse, CNRS, CNES, UPS, 14 Av. E. Belin, 31400 Toulouse, France,
\and 
Instituto de Astrof\'isica de Canarias, La Laguna
\and
LUPM, Universit\'e de Montpellier, CNRS, place Eugène Bataillon, 34095 Montpellier, France
\and
IRAP, Universit\'e de Toulouse, CNRS, UPS, CNES, 57 avenue d’Azereix, 65000 Tarbes, France
\and 
LESIA, Observatoire de Paris, Universit\'e PSL, CNRS, Sorbonne Universit\'e, Universit\'e Paris Cit\'e, 5 place Jules Janssen, 92195 Meudon, France
}

\date{Accepted in A\&A 16th May 2024}

\abstract
  {Both magnetic fields and photospheric/atmospheric dynamics can be involved in triggering the important mass loss observed in evolved cool stars. Previous works have revealed that these objects exhibit a magnetic field extending beyond their surface. The origin of this magnetic field is still under debate with mechanisms involving a turbulent dynamo, convection, stellar pulsation, and cool spots.}  
   {Our goal is to estimate the magnetic field strength in the inner circumstellar envelope of six evolved cool stars (five Miras and one Red Supergiant). Combining this work with previous studies, we tentatively constrain the global magnetic field type observed and shed light on the mechanisms at its origin.}
   {Using the XPOL polarimeter installed at the IRAM-30 m telescope, we observed the $^{28}\mathrm{SiO}$ v = 1, J = 2-1 maser line emission and obtained simultaneous spectroscopic measurements of the four Stokes parameters. Applying a careful calibration method for Stokes $Q$, $U$, and $V$, we derive estimates of the magnetic field strength from the circular and linear polarization fractions considering the saturated and unsaturated maser cases under the Zeeman hypothesis.  }
   {Magnetic field strengths from several Gauss up to several tens of Gauss are derived. These new and more accurate measurements constraining the field strength in the 2-5 stellar radii region better than previous studies and seem to exclude a global poloidal magnetic field type. A combination of a toroidal and a poloidal field is nevertheless not excluded. A variation of the magnetic field strength over a two-months timescale is observed in one Mira star which suggests a possible link to the stellar phase, i.e. with pulsation$/$photospheric activity. }
   {}
   \maketitle
%

\section{Introduction}
\label{sec:int}

After leaving the main sequence, solar-type stars (0.8-8 \Msol) will go through different phases prior to the planetary nebula stage \citep[e.g.][]{Herwig2005}. The Asymptotic Giant Branch (AGB) phase, even if lasting less than a few million years, is one of the most interesting in terms of chemistry and dynamics. AGB objects harbor in their circumstellar envelope (hereafter CSE), beyond the photosphere, a true chemical factory where more than 80 molecules and 15 dust species have been discovered \citep{hofner2018,Decin2018}. 
This AGB phase is, in addition, characterized by a massive mass loss \citep[$10^{-6}-10^{-4}$ \Msol$/$yr,][]{hofner2018} making these objects the main contributors to the chemical enrichment of the interstellar medium and the recycling of the matter in the Universe. This also applies to their massive counterparts, the Red Supergiants \citep[RSG,][]{Ekstrom2012,DeBeck2010}.

Evolved cool stars are hence providing strong mechanical and radiative feedback on their host environment \citep{Langer2012}. The high mass loss rates of AGB stars are thought to be the result of a wind acceleration mechanism based on radiation pressure on dust grains formed in the inner part of the CSE (at a few stellar radii) and  levitated by stellar pulsation \citep{Bladh2012}. For the RSGs no consistent scenario exists. The exact mechanisms that trigger and shape the strong winds of evolved cool stars still need further characterization. This requires deeper studies of sub-photospheric layers where convection and pulsation act, of the atmosphere where strong radiating shocks occur, and of the wind-forming region where dust condensates and radiation acceleration happens. In addition to the stellar convection, pulsation, and radiation pressure on dust, magnetic fields have been proposed to play a significant role through Alfv\'en-wave driving of the wind \citep[see for instance,][]{Cranmer, hofner2018}, but also in shaping the wind \citep[see][for pulsating evolved stars (AGBs) and for RSGs, respectvely]{Pascoli2008,Dorch2004}; if they are strong enough, they could even help in extracting the angular momentum \citep[cf. the case of the cool main sequence stars,][] {Bouvier2009}. Hence both magnetism and photospheric dynamics contribute in sustaining winds \citep{Lebre2014, Lebre2015, Lopez-Ariste2019}, but the relative importance of the magnetic field for the photospheric/atmospheric dynamics is still poorly known. 
Nevertheless, the knowledge of the magnetic field strength and geometry is still limited and its origin as a possible astrophysical dynamo in these stars would most likely be very different from the dynamo ($\alpha - \omega$) at work in solar-type stars, due to both their slow rotation and the fact that only a few convection cells are present at their surface at any given time \citep[][]{Freytag2002, Auriere2010}. For those slow rotators \citep[with high Rossby number, up to $\sim$ 100,][]{Charbonnel2017,Josselin2015}, classical $\alpha - \omega$ dynamos, where the toroidal component of the magnetic field is amplified by rotation, are indeed not expected. On the contrary, a turbulent$/$convective dynamo (${\alpha}^2 - \omega$) can generate a local weak magnetic field up to 0.01 Gauss which is then amplified by stellar pulsation and cool spots to a large-scale magnetic field \citep[][]{Soker2002}. In the case of RSGs, a local small-scale dynamo generating a strong field with a small filling factor is favored from numerical simulations \citep{Dorch2004}. 
Hence, while convection cannot generate global magnetic fields, local fields remain possible and may lead to the generation of (at least) local episodic mass loss events \citep[such as the one observed on Betelgeuse by][]{Montarges2021}. 

Low-intensity magnetic fields (of the order of 1-10 Gauss at the stellar surface or at a few stellar radii) have been identified and monitored over several years in RSGs and AGBs. Over the last decade, modern optical spectropolarimeters (CFHT/ESPaDOnS, TBL/Narval) have brought a lot of information on the stellar surface magnetism across the Hertzsprung-Russel diagram \citep[e.g.,][]{Auriere2010, Lebre2014, Tessore2017}.  In the radio regime, the polarimetric estimation of the magnetic field strength in the CSE of evolved stars has been possible thanks, for instance, to the XPOL instrument at the IRAM-30m \citep[e.g.,][]{Herpin2006}, the VLBA array \citep[e.g.,][]{Kemball2011, Assaf}, or now with ALMA \citep[e.g.,][]{Vlemmings2017}. From the radiation properties, such as linear polarization, the angle of polarization, and the circular polarization of maser emissions of different molecules, the magnetic field strength along the line of sight can be derived in the CSE of these objects \citep[see e.g. review of][]{Vlemmings2011}: from the innermost zones (i.e., a few stellar radii from the center of the object) via the SiO masers \citep{Herpin2006}, to the outermost layers (i.e., several thousand stellar radii) via OH masers for oxygen stars \citep{Rudnitski2010}, or via CN Zeeman effect for carbonaceous objects \citep{Duthu2017}. In particular, SiO masers are excited close to the star \citep[see e.g.][]{Cotton2011}, in small gas cells, where SiO has not yet been depleted onto grains at further distances \citep{Lucas1992,Sahai1993}; SiO and dust regions are sometimes overlapping in the near-CSE \citep{Wittkowski2007}. SiO masers exhibit linear and circular polarization which is likely due to the intrinsic stellar gas magnetic field \citep{Vlemmings2011}.  

Many results and positive detections have been obtained on the surface magnetic field from optical circular polarization of M spectral type AGB \citep[including the pulsating Mira variables,] []{Lebre2014, Konstantinova2014} and RSG stars \citep{Auriere2010, Tessore2017} revealing weak (i.e., down to the Gauss level) and variable or transient fields. From circular polarization in the radio regime, the magnetic field has been estimated at a few stellar radii in the CSE to be of the order of a few Gauss \citep[][]{Herpin2006}. All these observational constraints favor a magnetic field strength decreasing with a 1/r law throughout the CSE \citep{Duthu2017}. New insights came from ALMA interferometric observations of SiO linear polarization, revealing a magnetic field structure consistent with a toroidal field configuration \citep{Vlemmings2017}. Extrapolating this law towards the photosphere, for a toroidal magnetic field configuration \citep{Pascoli2008}, one can expect a magnetic field strength of a few Gauss at the surface of Mira stars. This was confirmed by the very first detection of a magnetic field at the surface of a Mira star, $\chi$ Cyg, \citep[2-3 Gauss,][]{Lebre2014}. The surface magnetic field appears to vary on time-scales of weeks to years, in agreement with the convective patterns timescale for RSGs \citep[e.g.,][] {Mathias2018}, or in the case of pulsating stars (Mira and RV Tauri stars) with the atmospheric dynamics, while the shock waves periodically propagating outward the stellar atmosphere may locally enhance the magnetic field through compression \citep[e.g.,] [] {Lebre2014, Lebre2015, Sabin2015, Georgiev2023}.

The origin of the observed surface magnetic field in those evolved cool stars, as well as the impact of the field strength on the stellar environment and subsequent evolution, remain to be fully characterized. Considering the difficulty to measure the magnetic field on the stellar surface of these objects, estimating its strength at different locations of the CSE, especially as close as possible to the photosphere, is a good way to advance our knowledge of the field configuration,
and therefore of its possible origin. In this paper, we present new SiO maser-line single-dish observations of six evolved cool stars, with improved calibration, in order to more accurately estimate the magnetic field strength in the inner region of the envelope, at 2-5 stellar radii where SiO masers are observed. In Sect. \ref{sec:obs}, we present the observations and simultaneous spectroscopic measurements of the four Stokes parameters. Section \ref{sec:datareduc} presents the data analysis and explains the calibration of our data, with emphasis on the removal of the instrumental polarization. The method that we used to derive the magnetic field estimates is presented in Sect. \ref{sec:datamethod}. In Sect. \ref{sec:results}, we present the polarization and magnetic field results. In Sect. \ref{sec:disc} we discuss the variability and origin of the magnetic field. The concluding remarks are developed in Sect. \ref{sec:Conclusions}. Further information is given in four Appendices.

\begin{table*}
\caption{List of the observed sources} 
\begin{center}
\label{tab:sources}
\begin{tabular}{cccccccccc}
    \hline
    \hline
    Star & RA (J2000) & DEC (J2000) & Type & d  & \vlsr & L  & \teff & dM/dt & $\phi$ \\
        &  (h m s) & ($^{\circ}$ ' '')  &  & (pc) & (\kms) & (\lsun) & (K) & (\Msol/yr) & \\ \hline 
     $o$ Ceti$^a$ & 02:19:20.79 & -02:58:39.50 & Mira (M5-9IIIe+DA) & 107 & +46.96 & 6099 & 2193& $2.5 \times 10^{-7}$ & 0.55\\
     R Leo$^b$ & 09:47:33.49 & +11:25:43.82 & Mira (M7-9eB) & 100 & 0.0 & 5617 & 2890 & $9.2 \times 10^{-7}$ & 0.85\\
     U Her$^{b,c}$ & 16:25:47.47 & 18:53:32.86 & Mira (M6.5-8+eB) & 271 & -16.48 & 8000 & 3000 & $5.9 \times 10^{-7}$ & 0.25/0.45\\
     R Aql$^{b,c}$ & 19:06:22.25 & 08:13:48.01 & Mira (M6-9eB) & 234/308 & +47.0 & 4900 & 2800 & $1.1 \times 10^{-6}$ & 0.90 \\
     $\chi$ Cyg$^b$ & 19:50:33.92 & +32:54:50.61 & Mira (S6-9/1-2eB) & 180 & +11.2 & 7813 & 2000 & $2.4 \times 10^{-7}$ & 0.90 \\
     $\mu$ Cep$^{d,e}$ & 21:43:30.46 & +58:46:48.16 & RSG (M2-IaB) & 390$/$1818 & +27.95 & 35000 & 3750 & $2.0 \times 10^{-6}$ & - \\
     \hline
\end{tabular}
\end{center}
\tablebib{
(a)~\citet{DeBeck2010}; (b) \citet{Andriantsaralaza}; (c) \citet{Gottlieb2022}; (d) \citet{Montarges2019};
(e) \citet{Maercker2022}.
}
\end{table*}

\section{Observations}
\label{sec:obs}
We observed the $^{28}\mathrm{SiO}$ v = 1, J = 2-1 maser-line emission at 86.2434277 GHz in a sample of evolved cool stars (see Table \ref{tab:sources}) in March 2022 and May 2022 at the IRAM-30m telescope on Pico Veleta, Spain. Simultaneous spectroscopic measurements of the four Stokes parameters $I$, $Q$, $U$, and $V$ have been obtained using the XPOL polarimeter \citep{Thum2008}. The EMIR front-end band E090 was connected to the VESPA backend, set up in polarimetry mode with a 120 MHz bandwidth and a channel separation of 40 kHz (i.e. $\sim$0.139 \kms ~at 86 GHz). EMIR (Eight MIxer Receiver) is one of the four dual-polarization heterodyne receivers available at the 30m facility (for more details, see the EMIR user guide\footnote{https://publicwiki.iram.es/EmirforAstronomers}). 
VESPA (The Versatile SPectrometer Array) is one of the seven available backends for EMIR\footnote{https://publicwiki.iram.es/Backends}. As an auto-correlation spectrometer, it was redesigned to also cross-correlate the orthogonal linear polarization signals recorded by EMIR, while it simultaneously delivers spectra in all four Stokes parameters\footnote{https://publicwiki.iram.es/PolarimetryforAstronomers}. The pointing was regularly checked on a nearby continuum source; accuracy is better than 3\arcsec \citep{Greve1996}. The focus was adjusted on an available planet. In order to obtain flat spectral baselines, we used the wobbler switching mode with a throw of 80\arcsec. The single sideband system temperature of the receiver was 105-110 K for both sessions. The integration times were 3.1, 2.0, 4.2, 3.2, 8.9, 1.9, and 0.2 hours for $\chi$ Cyg, $\mu$ Cep, $o$ Ceti, R Aql, R Leo, U Her (March), and U Her (May) leading to rms of 6.5, 8, 6, 7, 5, 7.5, and 24 mK, respectively, at the nominal spectral resolution. The forward and main-beam efficiencies were 0.95 and 0.81, respectively; the half-power beam width was 29\arcsec. The Jy/K conversion factor is 5.9.

Dedicated observations of the Crab Nebula (a well-characterized and strongly linearly polarized source) have been performed to verify the polarization angle calibration \citep[see][]{Thum2008} as well as planet observations with an unpolarized thermal emission (on Uranus) to estimate the instrumental polarization along the optical axis resulting from a feed leakage. The alignment between vertical and horizontal polarizations is perfect since the installation of an ortho-mode transducer with a single horn in 2016. The presence of polarized sidelobes does not affect our observations because our sources are not extended \citep[the SiO maser region, below typically 0.1 arcseconds in size, e.g.][is very small compared to the telescope beam]{Cotton2006}. The general methodology used for single dish SiO polarimetry
has been described in \citet{Herpin2006}. However, since 2015 the $V$ instrumental polarization has strongly increased due to a substantial leakage of the Stokes $I$ signal into the Stokes $V$ \citep[][]{Duthu2017}. An improved calibration scheme has then been used in the present work to minimize this contamination and is presented in Sect.\ref{sec:instr}.

Our sample is shown in Table \ref{tab:sources} (the stellar phase $\phi$ is given in column 10) and consists of five Mira-type stars (R Aql, $o$ Ceti, $\chi$ Cyg, U Her, and R Leo) and one RSG ($\mu$ Cep). All sources are known to exhibit strong SiO maser emission. R Leo and $\chi$ Cyg have already been observed by \citet{Herpin2006} in polarimetric mode. Note that U Her was observed in both March and May 2022.

\label{tab:obs}

\section{Data reduction and calibration}
\label{sec:datareduc}
\subsection{Stokes parameters}
\label{sec:stokes}

From the horizontal and vertical polarizations and their relative phase shift, we can simultaneously determine the four Stokes parameters:

\begin{equation}
\begin{array}{lcl}
    I &=&\langle E_{\textrm{H}}^2 \rangle + \langle E_{\textrm{V}}^2 \rangle\\ [0.5ex]
    Q &=& \langle E_{\textrm{H}}^2\rangle - \langle E_{\textrm{V}}^2 \rangle \\ [0.5ex]
    U &=& 2\langle E_{\textrm{H}}E_{\textrm{V}}\textrm{cos} \delta \rangle\\ [0.5ex]
    V &=& 2 \langle E_{\textrm{H}}E_{\textrm{V}}\textrm{sin} \delta \rangle
    \end{array}
\end{equation}

with $\delta$ the phase difference between horizontal and vertical components $E_{\textrm{H}}$ and $E_{\textrm{V}}$ \citep[e.g.][]{Landi2004}. 

\begin{figure}
    \includegraphics[width=0.9\columnwidth]{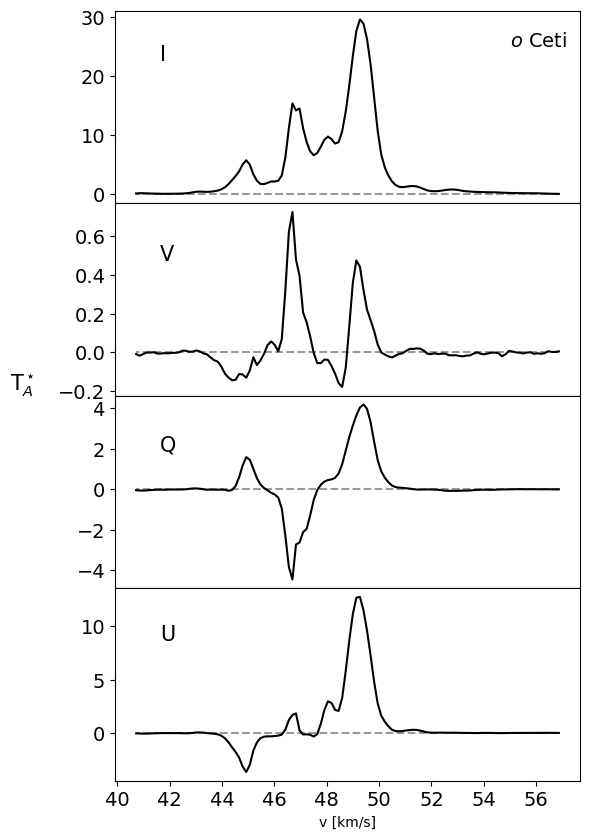}
    \caption{Observed Stokes parameters (not corrected for instrumental polarization) $I$, $V$, $Q$, and $U$ for $o$ Ceti (in T$_A^\star$ units [K]), spectral resolution of 0.139 \kms \,centered on the SiO transition.}
    \label{fig:Stokes} 
\end{figure}

As polarizations are measured in the Nasmyth cabin reference frame, Stokes parameters $Q$ and $U$ are then rotated to the equatorial system to correct for the parallactic rotation. The calibration of the phase difference between the orthogonally polarized signals is automatically determined during the observations, by injecting a signal from a liquid-nitrogen bath passing through a well-specified wire grid \citep{Thum2008}, and applied by dedicated offline data-processing. An example of $I, Q, U, V$ spectra is shown in Fig. \ref{fig:Stokes} for $o$ Ceti.

The data reduction was made with the CLASS\footnote{http://www.iram.fr/IRAMFR/GILDAS/} package. The first step consists of removing linear baselines from all individual spectra. In a second step, we correct the measured polarizations for the instrumental leakage from Stokes $I$ to $Q$, $U$, and $V$; the used methodology is described in Sect. \ref{sec:instr}. The third step consists of averaging, under sensitivity-optimizing weighting, all spectra for a given Stokes parameter to form a single spectrum to be analyzed.

From those, we can compute the fractional circular polarization ($p_\textrm{C}$), the fractional linear polarization ($p_\textrm{L}$), and the angle of polarization ($\chi)$. These are defined as \citep{Landi2004}:

\begin{equation}
\begin{array}{lcl}
    p_\textrm{C} &=& V/I\\ [0.5ex]
    p_\textrm{L} &=& \sqrt{Q^2 + U^2} / I \\ [0.5ex]
    \chi &=& \frac{1}{2} \textrm{atan2}(U,Q) \,= \frac{1}{2}\mathrm{Arg}(U+iQ)
\label{eq:param}
\end{array}
\end{equation}

The astronomical signal that we get is an incoherent mixture of Stokes $Q$, $U$, and $V$ parameters filling the observing beam, i.e., originating from slightly different sight-lines. 
The use of the four-quadrant inverse tangent function instead of the arctangent function avoids $90^\circ$ ambiguities arising when a significant polarization in Stokes $U$ is measured together with an insignificant Stokes $Q$ value of undefined sign. Owing to the modulation of Stokes $Q$ and $U$ in a non-corotating Nasmyth reference frame, such a situation is unavoidable and frequent.

\subsection{Instrumental Polarization} 
\label{sec:instr}

Most astronomical telescopes and their instrumentation suffer from spurious conversions from Stokes $I$ into Stokes $Q$, $U$, and $V$, and among the three latter. The reasons for these conversions are manifold and mainly originate in the asymmetric optics of the telescope (e.g., Nasmyth designs like that of the IRAM-30m) and the re-imaging system of the receivers, and in the receiver feed horns. For a discussion, we refer to
\citet[][and further references therein]{Thum2008}, demonstrating the extremely low instrumental conversion of the IRAM-30m optics from Stokes $I$ into Stokes $V$ (Müller matrix element $M_\mathrm{IV}$ at the per-mil level even outside the optical axis, after optimization of the orientation of the beam-splitting grid). In December 2015, the E090 band of EMIR was equipped with an ortho-mode transducer splitting the orthogonal linear polarizations only in the feed horn. Its imperfections induce an instrumental polarization of up to $M_\mathrm{IV} = 2.7$\%, derived from observations of the unpolarized planet Uranus. Calibration of the polarization angle is as good as $\pm 0\fdg 5$, which is the uncertainty of the orientation of the wire grid in the phase-calibration unit. A spurious conversion from Stokes $U$ to $V$ may arise in the phase calibration, but is shown to fall below $M_\mathrm{UV} \la 1$\%. 
From the Crab Nebula observations performed at the beginning and end of each period, we derive a stable polarization angle of 145-146$^{\circ}$ in agreement with what is expected \citep{Aumont2010,Ritacco2018}. 
This demonstrates that the instrumental conversions between linear and circular polarization are under control. For brevity, we
therefore use \textit{instrumental polarization} as a synonym for the leakages from Stokes $I$ into $Q$, $U$, and $V$, referring to specific conversions with the Müller matrix elements $M_\mathrm{IQ}$, $M_\mathrm{IU}$, and $M_\mathrm{IV}$. In the following, we describe the method used to determine the instrumental polarization and to remove it from the Stokes spectra. It is important to note that these calibration steps are applied scan-wise, i.e., to spectra of typically four-minute long on-off cycles. This is because the instrumental polarization arises predominantly in the Nasmyth cabin and, therefore, if expressed in celestial reference frames depends on elevation. Final spectra are subsequently obtained by averaging the calibrated scans.

\begin{table}[]
    \centering
    \caption{Global uncertainties (standard deviations) of Stokes $V$ derived from the Monte-Carlo method, including uncertainties in the removal of instrumental contamination.}
    \begin{tabular}{lcc}
         \hline
         \hline
         &  $\sigma_V$ & Integration time\\
         & (mK) & (hour)\\
         \hline
         $o$ Ceti & 1.9 & 4.2\\
         R Leo & 1.1 & 8.9 \\
         U Her (March) & 2.4 & 1.9 \\
         U Her (May) & 6.9 & 0.2 \\
         R Aql & 5.1 & 3.2\\
         $\chi$ Cyg & 3.8 & 3.1 \\
         $\mu$ Cep & 4.8 & 2.0 \\
    \end{tabular}
    \label{tab:MC}
\end{table}

\subsubsection{Removal of instrumental leakage from Stokes $I$ to Stokes $V$} 
\label{sec:instr_I_V}
A given Stokes $V$ spectrum is composed of an astronomical signal, $V^\star$, and an instrumental leakage from Stokes I, i.e.,
$V = V^\star+M_\mathrm{IV} \cdot I$. Our recovery of $V^\star$ then proceeds as follows:
\begin{enumerate}
    \item Since the $M_\mathrm{IV}$ term is known to be quasi-achromatic, it can be quantified, for each scan, by the linear regression of Stokes $I$ and $V$. By averaging only spectra with $|M_\mathrm{IV}| < 1$\%, we obtain an almost pure, preliminary astronomical Stokes $V$ spectrum, $\tilde{V}^\star$.
    \item $\tilde{V}^\star$ is subtracted from each Stokes $V$ spectrum. The resulting scan series of residual spectra $\Delta V$ is then largely dominated by the leakage terms $M_\mathrm{IV}\cdot I$. Estimates of $M_\mathrm{IV}$ are scan-wise determined by linear regression of Stokes $I$ and $\Delta V$.
    \item In the final step, we compute $V^\star = V-M_\mathrm{IV}I$. The thus obtained series of corrected spectra with baseline noise $\sigma_\mathrm{rms}$ are finally averaged with a sensitivity-optimizing $1/\sigma_\mathrm{rms}^2$ weighting.
\end{enumerate}

\noindent A demonstration of this algorithm is shown in Fig.~\ref{fig:Appen_ipv}.
\begin{figure*}
    \centering
    \includegraphics[width=0.45\textwidth]{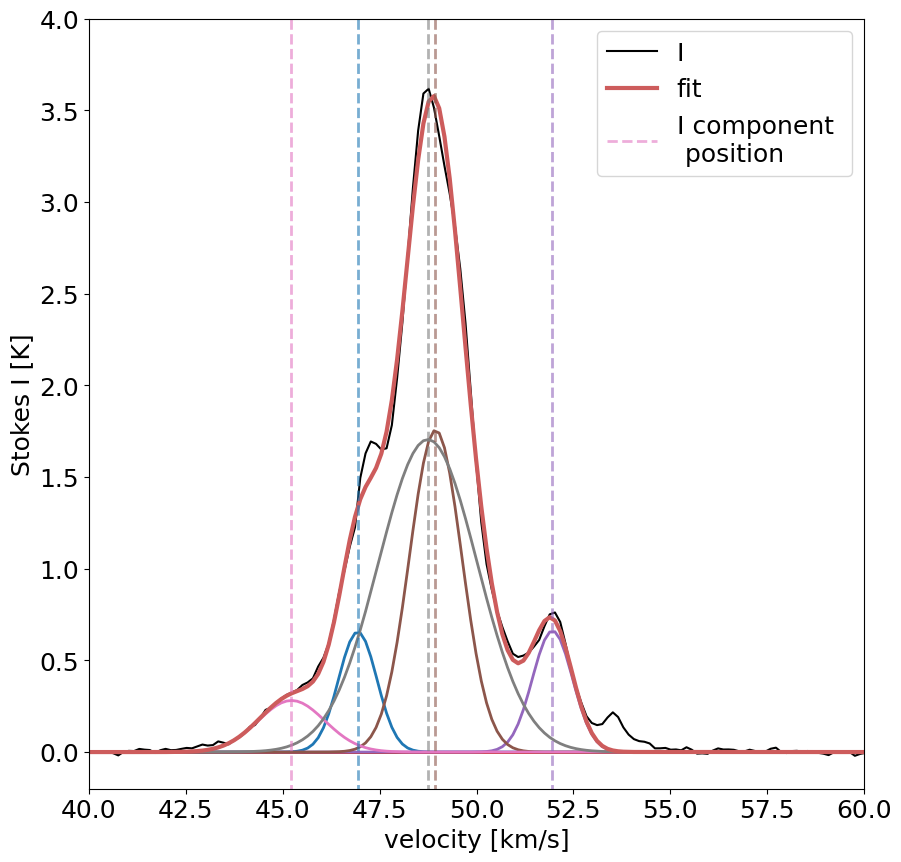}
    \includegraphics[width=0.47\textwidth]{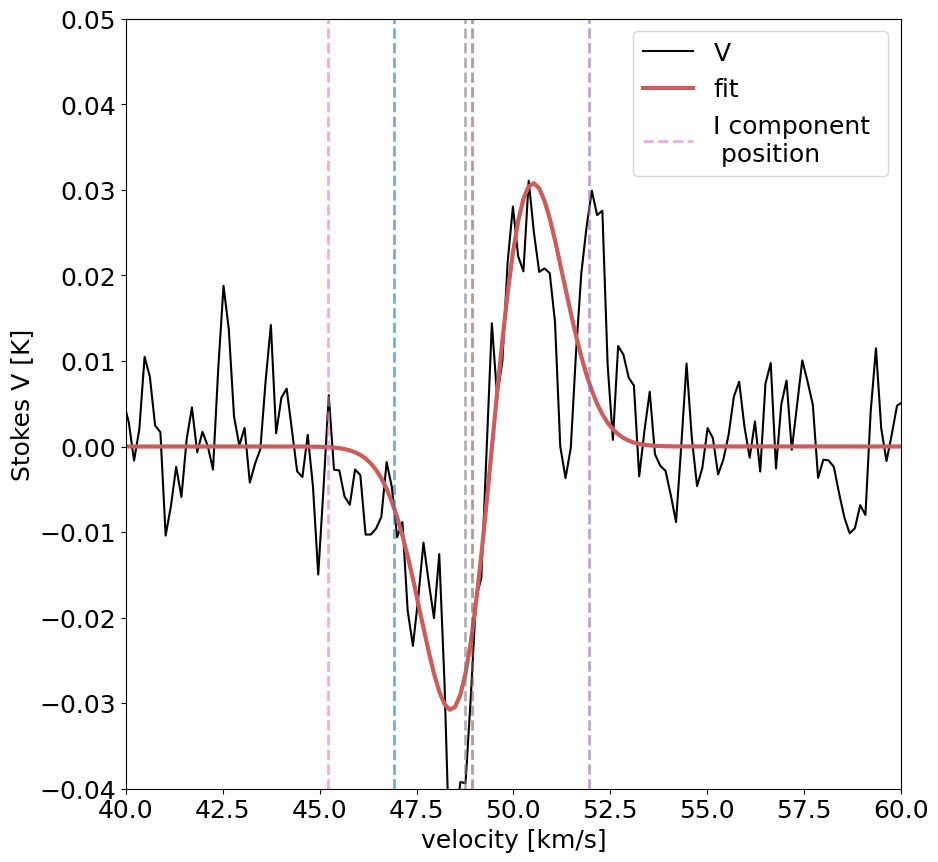}
    \caption{Decomposition of Stokes $I$ and $V$ for R Aql (spectral resolution of 0.139 \kms). \textbf{Left}: In black the Stokes $I$ data, in red the global fit; the vertical colored dashed lines are the positions of the center of the individual Gaussian functions used in the fit, plotted in various colors. \textbf{Right}: In black, the Stokes $V$ data cleaned from instrumental polarization (Sect. \ref{sec:instr}), the fit is in red; the vertical dashed colored lines show the individual Gaussian $I$ components from Stokes $I$ fitting procedure (left panel).}
    \label{fig:raqldecompo}
\end{figure*}
The residual instrumental polarization in the cleaned Stokes $V$ spectra mainly arises in the uncertainties $\Delta M_\mathrm{IV}$ of the thus obtained parameters $M_\mathrm{IV}$ and is estimated by applying the same calibration steps to synthetic Stokes $V$ spectra of an underlying $V^\star$ known a priori:
\begin{enumerate}
    \item As data model for a given target, we use its $V_\mathrm{rvm}^\star$ spectrum produced by the \textit{rvm} code (see Sect. \ref{sec:StokesDec}). For each scan out of the total of $N_\mathrm{obs}$ observations we add baseline noise of standard deviation $\sigma_\mathrm{rms}$ and an instrumental contribution $(M_\mathrm{IV}+\Delta M_\mathrm{IV})I$. The $N_\mathrm{obs}$ realizations of the synthetic noise spectra and $\Delta M_\mathrm{IV}$ parameters are obtained from Gaussian distributions of the same standard deviations as those determined in the calibration steps described above.
    \item We decontaminate this set of $N_\mathrm{obs}$ simulated spectra and deduce the best estimate $\hat{V}^\star$ of the astronomical Stokes $V$ spectrum, using the same cleaning procedure as described above.
    \item We repeat this procedure ten times to obtain for each velocity channel the standard deviation among the $\hat{V}^\star$ spectra, and assign it 
    to the corresponding channel of the $V_\mathrm{rvm}^\star$ spectrum.
\end{enumerate}
For all sources, we apply this Monte-Carlo method and we obtain standard deviations $\sigma_\mathrm{V}$ of the order of a few mK, reported in Table \ref{tab:MC}. They exceed those of the radiometric noise in the original Stokes $V$ spectra and reflect the residual uncertainty in the determination of the $M_\mathrm{IV}$ parameters.
We underline that for R Aql, the previous test for the reproducibility of the required correction indicates that the derived $B\textsubscript{Z} \times \textrm{cos}({\theta}_2)$ is inconclusive. It is only formally significant the $V_\mathrm{rvm}^\star$ spectrum uncertainty is only based on the radiometric noise; there are no limitations linked to the procedure removing the $M_\mathrm{IV}$ terms.

\subsubsection{Removal of instrumental leakage from Stokes $I$ to $Q$ and $U$}
\label{sec:instrUQ}

The removal of the leakage from Stokes $I$ to linear polarization is conceptually easier because in the celestial reference frame \citep[where the polarization angle is measured in the equatorial system from the north towards the east following the IAU convention,][]{IAU73} the astronomical polarization explicitly depends on the parallactic rotation, or, to be more precise, on the difference $\chi_0$ between the telescope's elevation $\varepsilon$ and the parallactic angle $\eta$, whereas the instrumental contribution remains fixed in the reference frame of the Nasmyth cabin. For the EMIR receiver, the transformation from celestial coordinate offsets 
defined in the equatorial frame $(\Delta\alpha\cos{\delta},\Delta\delta)$ to those defined in the Nasmyth frame is given by
\begin{equation}
    \mathbf{T} = \begin{pmatrix}
                           \cos\chi_0 & \sin\chi_0 \\
                          -\sin\chi_0 & \cos\chi_0
                      \end{pmatrix}.
\end{equation}
The application of the coherency matrix formalism \citep[e.g.,][further references therein]{Wiesemeyer14} then yields
the corresponding transformation of the equatorial plane of the Poincar\'e sphere to the Nasmyth reference frame (subscript $N$), with
the following representation of the Stokes $Q$ and $U$ parameters:
\begin{equation}
   \begin{matrix}
      Q_\mathrm{N} & = & M_\mathrm{IQ}I & + & Q^*_\mathrm{eq}\cos{2\chi_0} & + & U^*_\mathrm{eq} \sin{2\chi_0}, \\
      U_\mathrm{N} & = & M_\mathrm{IU}I & - & Q^*_\mathrm{eq}\sin{2\chi_0} & + & U^*_\mathrm{eq} \cos{2\chi_0},
   \end{matrix}
   \label{eq:transform_qu}
\end{equation}
where the superscripts $^*$ refer as in Sect. \ref{sec:instr_I_V} to the astronomical Stokes parameters, and where $M_\mathrm{IQ}$ and $M_\mathrm{IU}$ are the M\"uller matrix elements for the spurious conversion of Stokes $I$ into $Q$ and $U$. The knowledge of the exact nature of the $M_\mathrm{IQ}$ and $M_\mathrm{IU}$ terms is not required here, except that their origin is located in the Nasmyth cabin, and that they are quasi-achromatic across the lines profiles here. From previous commissioning work \citep{Thum2008}, these prerequisites are known to be fulfilled. Under the well-justified premise that the emission in the Stokes parameters $Q^*$ and $U^*$ is spatially unresolved and stationary on the timescale of an hour-angle interval, and ensuring sufficient sampling of $2\chi_0$, they can be separated from the instrumental contamination by fitting sinusoidal curves to the line-integrated fluxes using Eq.~\ref{eq:transform_qu}. An example is shown in Fig.~\ref{fig:Appen_ipqu}. Minor, systematic deviations hint at another instrumental contribution to the measured polarization that is fixed in the horizontal system and therefore arises from the telescope's primary or secondary mirror. These contributions were found to be insignificant.
%

\section{Data analysis and method}
\label{sec:datamethod}
\subsection{Stokes $I$ and $V$ components analysis}
\label{sec:StokesDec}

The SiO Stokes $I$ spectra are shown in Fig. \ref{fig:raqldecompo} for R Aql and in Appendix B (Fig. \ref{fig:stokesI}) for the other sources. All SiO line profiles are complex, but can be reproduced by assembling several Gaussian maser components centered at different velocities. The individual components are relatively narrow, although some of them are slightly broader than the thermal linewidths, of the order of 1 to 1.2 \kms \, in SiO maser regions where we expect the gas temperature to be around 1000 to 1500 K. 
We have decomposed the Stokes $I$ signal into multiple Gaussian functions, assuming that the intensity profile $I$ can hence be fitted by:

\begin{equation}
    I(\varv) =  \sum_{k=1}^n A_k e^{-\frac{(\varv-\varv_{0,k})^2}{\Delta \varv_{D,k}^2}}
    \label{eq:I}
\end{equation}
with $\varv_{0,k}$ and $\Delta \varv_{D,k}$ respectively the central velocity and Doppler width for each component $k$, and $A_k$ a weight coefficient. 

We have developed a code based on the Relevance Vector Machines (rvm) algorithm as described in \cite{Tipping}. This code fits the signal to the best linear combination of entries taken from a dictionary. In our case, the dictionary is made of Gaussian functions of various widths (Full Width at Half Maximum, FWHM), that are chosen in a range of two values, $\Delta \varv_a$ and $\Delta \varv_b$, sampled in 50 values. They are defined for each star and centered over a range of velocities (200 values between $\varv_0$ and $\varv_1$). The term \textit{dictionary} is used to emphasize that there is no requirement of orthogonality among the different elements of the dictionary. Lacking such a requirement, the algebraic problem has no unique solution. \textit{rvm} solves this issue by Bayesian marginalization. The code assigns a weight to all the components of the dictionary ($A_k$), optimizing the marginal likelihood. In a sense, the algorithm seeks the linear combination with the smallest number of dictionary entries which satisfies noise constraints, consistent with the noise observed in our data.

Initial fits were made with the CLASS package to get a first idea of ranges for widths and velocities. Those parameters were then used as a first guess for the ranges in the \textit{rvm} code. The ranges were adjusted iteratively to maximize the correlation between the fit and the observed spectra. 
he results are not sensitive to the adopted ranges provided that they remain within reasonable values. The FWHM ranges differ slightly for each source, but they typically are between 0.3 and 2 \kms. The fit result is more sensitive to the noise level put in the algorithm, i.e. the Stokes \textit{rms}, that will constrain the number of components used in the fit.

\begin{figure*}
    \centering
    \includegraphics[width=0.95\textwidth]{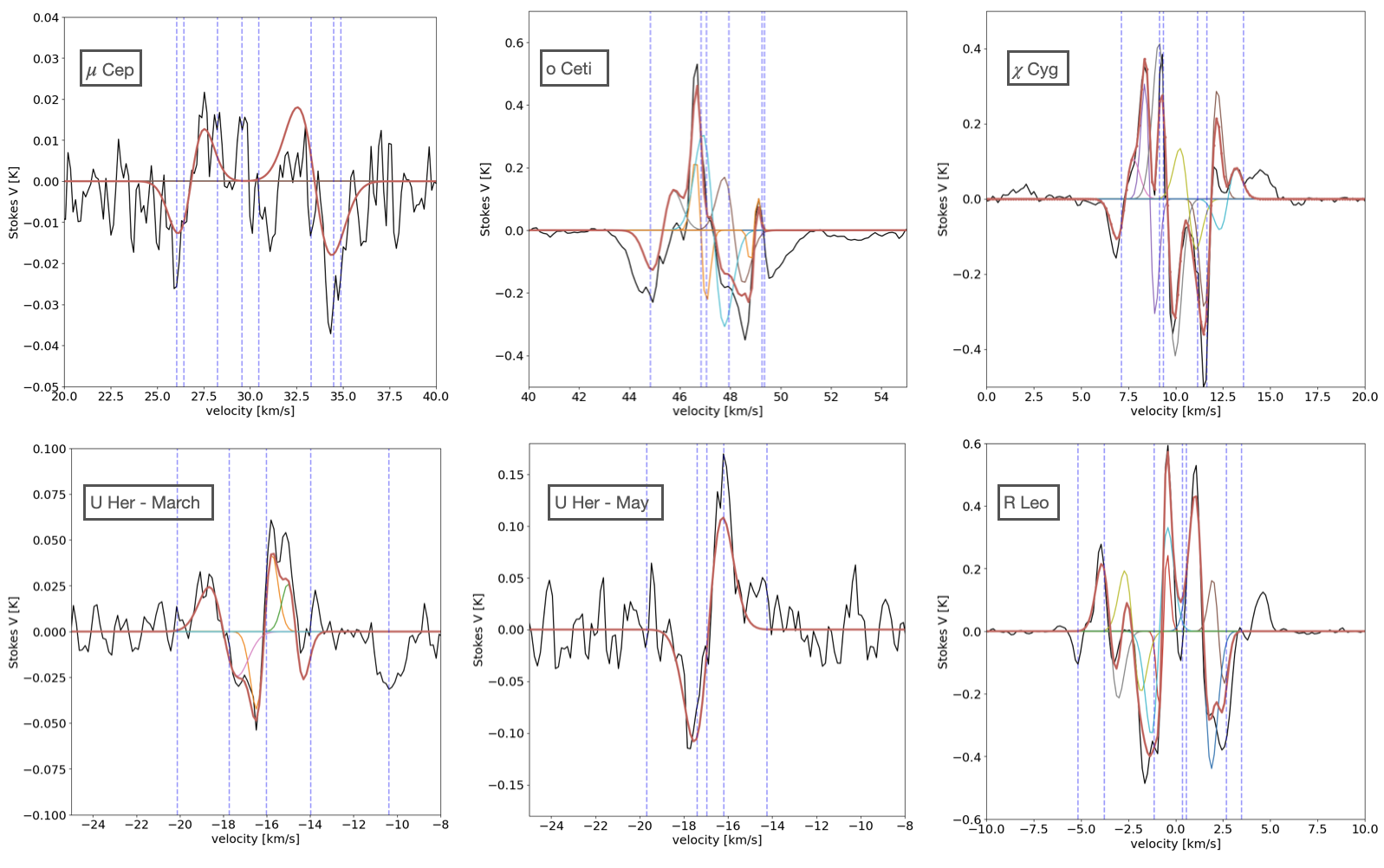}
    \caption{Stokes $V$ spectra for all sources, except R Aql. In black, the Stokes $V$ data cleaned from instrumental polarization (Sect. \ref{sec:instr}), the fit in red; the dashed purple lines are the positions of the individual Gaussian $I$ components from Stokes $I$ fitting procedure.}
    \label{fig:V_fitted}
\end{figure*}

Always assuming that circular polarization has its origin in the Zeeman effect, an analogous decomposition is used for Stokes $V$ profiles. Since SiO is a non-paramagnetic molecule, the typical magnetic fields expected to be found in the CSE correspond to the so-called weak-splitting Zeeman regime, for which Zeeman splitting is much smaller than the intrinsic Doppler line width  $\Delta \varv_{D}$. This is independent of any line width growing in the maser theory, but we will describe later the impact of maser saturation. In the weak field regime, any Stokes $V$ signal can be safely modelled: 
\begin{equation}
    V(\varv) = f(B) \frac{dI}{d\varv}
    \label{eq:Vrvm}
\end{equation}
where $f(B)$ is a function of the magnetic field intensity and the parameters of the transition, with respect to the associated intensity. As in the case of the observed Stokes parameter $I$, the observed parameter $V$ is broken down into subcomponents $V_i$ that are derivative of Gaussian functions which, when added together, fit the $V$ profile. Our \textit{rvm} algorithm determines the weights assigned to all $V_i$ whose weighted combination fits the observed $V$. All profiles are treated here as one would do with thermal Zeeman measurements, but of course maser amplification may modify the line profiles (maser saturation will be considered in Sect. \ref{sec:saturation}).

Fit results on Stokes $I$ and $V$ are given in Tables \ref{tab:I}-\ref{tab:V} and are shown in Fig. \ref{fig:raqldecompo} for R Aql and in Fig. \ref{fig:stokesI} (Stokes $I$) and \ref{fig:V_fitted} (Stokes $V$) for the other sources.

\subsection{Computation of uncertainties}
\label{sec:uncert}

To compute the uncertainties $\delta B$ of the inferred magnetic field, we need to compute the uncertainties $\delta V$ of the Stokes $V$ parameters after the removal of the instrumental polarization as described above. We consider two main sources of uncertainties: 

\begin{itemize}
    \item 
    The Monte-Carlo method described in \ref{sec:instr_I_V} includes the Stokes $V$ global uncertainty for each velocity channel, i.e., reflects both the radiometric noise and the uncertainty in the removal of the $M_\mathrm{IV}$ terms
    (see $\sigma_V$ in Tab. \ref{tab:MC}). 
    This entails an uncertainty $\delta V_{\textrm{inst}}$ in the \textit{rvm} fit which needs to be quantified.
    \item The difference ("residual"), $\delta V_{\textrm{res}}$, between the fit and the cleaned spectra $V^\star$ (see Sec. \ref{sec:instr_I_V}).
\end{itemize}

We can quantify the first uncertainty from the standard deviations $\sigma_\mathrm{V}$ as determined in Sect. \ref{sec:instr_I_V}, which propagates into the parameters of the Stokes $V$ fitting method and, consequently, to the magnetic field estimates.
To determine this first uncertainty, we define three spectra, namely the cleaning-method spectrum and two spectra deviating from the cleaned one by $\pm \sigma_\mathrm{V}$.
%
We apply the \textit{rvm} code to these three spectra to see if there are differences in the parameters of the Gaussian functions (velocity widths, central velocities, and amplitudes) used in the fit. We find a difference of 0.3 \kms \, in the central velocity for U Her in May and 0.1 \kms \, for the three sources $\mu$ Cep, $o$ Ceti, and U Her (March), while there is no difference for the other objects. We also determine the corresponding difference $\delta V_{\textrm{inst}}$ between the largest excursions introduced in the peak intensities $V_\textrm{peak}$; it amounts to 7~mK. 

To estimate $\delta V_{\textrm{res}}$, we measure the \textit{Root Mean Square Error} ($\delta V_{\textrm{res}} = \sqrt{\sum_{i=1}^{N} (x_i^m - x_i^d)^2 /N}$ with $x^m$ the fitted values and $x^d$ the observational data). For R Aql, $\mu$ Cep, and U Her (March observation) we find $\delta V_{\textrm{res}} \sim 7$ mK and for U~Her (May observation), R~Leo, $\omicron$~Ceti and $\chi$~Cyg 24, 28, 28 and 33 mK, respectively. 

To obtain the total uncertainty $\delta V$ in Stokes $V$ we add $\delta V_\mathrm{inst}$ and $\delta V_\mathrm{res}$ in quadrature; the results are reported in Table \ref{tab:V}. Thanks to their large signal-to-noise ratios, we assume the Stokes $I$ spectra to be free from errors.
For both regimes (i.e., saturated and unsaturated), 
uncertainties in the magnetic field estimates
are then given by $\delta B / B = \delta V / V$.

\subsection{Description of the observed Stokes $I$ and $V$ signals}
\label{sec:StokesI}

\begin{table*}[h!]
\centering
\caption{Description of the $I$ components for each source and associated saturation rates.} 
\label{tab:I}
\begin{tabular}{ccccccccc}
    \hline
    \hline
    Star& rms & v & Flux & $\Delta$v & S$/$N & T$_B$ & R  & R$/\Gamma$\\
    & (mJy) & (\kms) & (Jy) & (\kms) && ($10^7$ K) & (s$^{-1}$) &\\
    \hline 
    R Aql & 43 & 45.2 & 1.7 & 1.2 & 14 & 0.8$-$1.4 & 0.2$-$0.4 & 0.04-0.08\\  
&& 46.9 & 3.9 & 0.7 & 32 & 1.7$-$3.1 & 0.5$-$0.9 & 0.1$-$0.2 \\  
&& 48.7 & 10.1 & 1.8 & 83 & 4.5$-$8.1 & 1.3$-$2.3 & 0.3$-$0.5 \\  
&& 48.9 & 10.4 & 1.0 & 85 & 4.6$-$8.4 & 1.3$-$2.3 &0.3$-$0.5  \\  
&& 52.0 & 3.9 & 0.7 & 32 & 1.7$-$3.1 & 0.5$-$0.9 & 0.1$-$0.2\\ 
    $\mu$ Cep& 46 & 26.0 & 3.3 & 0.4 & 25.0 & 4.1$-$89.9 & 1.1$-$25 & 0.2$-$5 \\  
&& 26.4 & 4.0 & 1.4 & 31 & 5.0$-$109.0 & 1.4$-$30.3  & 0.3$-$6.1\\  
&& 28.2 & 6.0 & 1.5 & 46 & 7.5$-$163.4 & 2.1$-$45.4 & 0.4$-$9.1 \\  
&& 29.5 & 2.1 & 0.3 & 16 & 2.6$-$57.2 & 0.7$-$15.9 & 0.1$-$3.2\\  
&& 30.5 & 4.3 & 1.1 & 33 & 5.4$-$117.1 & 1.5$-$32.5 & 0.3$-$6.5\\  
&& 33.3 & 1.9 & 0.3 & 15 & 2.4$-$51.8 & 0.7$-$14.4 & 0.1$-$2.9\\  
&& 34.5 & 4.6 & 1.4 & 35 & 5.8$-$125.3 & 1.6$-$34.8 & 0.3$-$7.0\\  
&& 34.9 & 2.2 & 0.5 & 17 & 2.8$-$59.9 & 0.8$-$16.6 & 0.2$-$3.3\\
    $o$ Ceti & 36 & 44.8 & 26.2 & 0.6 & 260 & 2.5 & 0.7 & 0.1 \\  
&& 46.8 & 68.7 & 0.4 & 682 & 6.5 & 1.8 & 0.4 \\  
&& 47.0 & 26.5 & 1.0 & 263 & 2.5 & 0.7 & 0.1 \\  
&& 47.9 & 37.4 & 0.4 & 372 & 3.6 & 1.0 & 0.2 \\  
&& 49.2 & 93.8 & 0.9 & 932 & 8.9 & 2.5 & 0.5\\  
&& 49.3 & 93.7 & 0.5 & 931 & 8.9 & 2.5 & 0.5 \\
     $\chi$ Cyg & 38 & 7.1 & 39.4 & 1.2 & 366 & 10.5 & 2.9 & 0.6\\  
&& 9.1 & 290.2 & 0.5 & 2693 & 77.6 & 21.6& 4.3 \\  
&& 9.3 & 172.7 & 1.1 & 1603 & 46.2 & 12.8 & 2.6 \\  
&& 11.2 & 64.2 & 1.2 & 596 & 17.2 & 4.9 &  1.0 \\  
&& 11.7 & 124.3 & 0.5 & 1154 & 33.2 & 9.3 & 1.9 \\  
&& 13.6 & 15.1 & 1.2 & 140 & 4.0 & 1.1 & 0.2 \\  
    U Her & 45 & -20.1 & 3.7 & 1.3 & 29 & 2.2 & 0.6 & 0.1 \\  
(March)&& -17.7 & 30.2 & 1.0 & 240 & 18.3 & 5.1 & 1.0 \\  
&& -16.0 & 31.9 & 1.0 & 254 & 19.3 & 5.4 & 1.1 \\  
&& -14.0 & 6.9 & 1.3 & 55 & 4.2 & 1.2 & 0.2 \\  
&& -10.4 & 2.1 & 2.0 & 17 & 1.3& 0.4 & 0.1 \\  
    U Her & 140 & -19.7 & 2.9 & 0.9 & 73 & 1.8& 0.5& 0.1 \\  
(May)&& -17.4 & 29.0 & 0.6 & 731 & 17.5 & 4.9 & 1.0 \\  
&& -17.0 & 32.9 & 1.3 & 830 & 19.9& 5.5 & 1.1 \\  
&& -16.2 & 19.6 & 0.6 & 495 &11.9 & 3.3& 0.7 \\  
&& -14.2 & 3.2 & 1.5 & 81 & 1.0 & 0.3 & 0.1 \\ 
    R Leo & 29 & -5.2 & 15.8 & 0.7 & 196 & 1.3 & 0.4 & 0.1 \\  
&& -3.8 & 47.2 & 0.5 & 587 & 3.9 & 1.1 & 0.2 \\  
&& -1.2 & 137.6 & 0.7 & 1710 & 11.3 & 3.1 & 0.6 \\  
&& -0.2 & 162.3 & 2.3 & 2018 & 13.4 & 3.7 & 0.7 \\  
&& 0.6 & 117.9 & 0.8 & 1465 & 9.7 & 2.7 & 0.5 \\  
&& 2.7 & 21.6 & 0.4 & 268 & 1.8& 0.5 & 0.1 \\  
&& 3.5 & 51.3 & 1.5 & 637 & 4.2 & 1.2 & 0.2 \\  
    \hline 
\end{tabular}
\tablefoot{We considered for $\mu$ Cep and R Aql two different distances (see Table \ref{tab:sources})}
\end{table*}

Our fit results (see Figs. \ref{fig:raqldecompo} for R Aql, \ref{fig:V_fitted} for the $V$ decomposition, \ref{fig:stokesI} for $I$, and Tables \ref{tab:I}, \ref{tab:V}) show that the SiO maser line profiles are globally well reproduced by our procedure. Exceptions can be found in the bluest and reddest weak features of the total intensity $I$, as for instance at $\sim$54 \kms \; for R Aql, but also for $\chi$ Cyg, $o$ Ceti, or U Her (for the May data). Some broad wing emissions in $\mu$ Cep, $\chi$ Cyg, or $o$ Ceti (including component \#1) are not reproduced either. This has no impact on our results as the polarization of these components is weak, within the noise. All detected Stokes $I$ components have amplitudes well above 10$\sigma$ and are $\sim$1 \kms \; (or less) broad, as expected from maser emission \citep[e.g.][]{Richards2020}. However, one central component in R Leo has a width of 2.3 \kms~but might be due to an underestimation (too weak) of the other components on both sides. We underline that the maser emission in U Her has changed between the two epochs of observation. For this source, the profiles are different, although we identify the same number of components (five) but at different velocities. Indeed, the last component at -10.4 \kms \, in March has no equivalence in May. The first components at -20.1 \kms \,(March) and -19.7 \kms \,(May) have the same intensity and width. Between the components at -17.7/-17.4 \kms~ and -16.0/-16.2 \. \kms, which are weaker in May, the code has found for May an additional, larger ($\Delta \varv = 1.3$ \kms), component at -17.0 \kms \,(See Figure \ref{fig:stokesI}).

Two types of $V$ profiles are globally observed within our sample: single profiles with 1-2 well-separated Zeeman components (R Aql, U Her in May, and $\mu$ Cep) and more complex profiles with more components, some of them overlapping. U Her in March is between the two cases and does not exhibit the simple anti-symmetric profile observed in May. It is important to note that not all $I$ components are associated with characteristic anti-symmetric Zeeman $V$ profiles; this could perhaps be due to the complexity of the magnetic field within our rather large radio beamwidth. The Zeeman features in Stokes $V$ that can be associated with a Stokes $I$ component are (blue- or red-) shifted by a few 0.1 \kms, hence in general more than our uncertainty on the fits (see Sect. \ref{sec:uncert}) and comparable to or more than the channel separation. This velocity shift is thus real, but we cannot explain it with a simple model associating physical maser components with emission peaks and a magnetic field. We cannot exclude from the present work that other theories without standard Zeeman interpretation could lead to complex profiles with possible velocity shifts.

The $V$ spectra for R Aql and U Her (May data) exhibit a well-defined single Zeeman signature, red-shifted from the corresponding $I$ component. $\mu$ Cep exhibits two well-separated Zeeman signatures, also red-shifted, but with a low S$/$N. The fit for U Her (March data) indicates three components, but considering the S$/$N, only the central one (in orange in Figure \ref{fig:V_fitted}), roughly identical to that detected in May, can be trusted. The $V$ profiles for $\chi$ Cyg, R Leo, and $o$ Ceti are made of 5 or 6 components each, most of them associated with $I$ components though with noticeable velocity shifts. $o$ Ceti exhibits broad negative  $V$ signals in the  bluest and reddest sides of the $V$ profiles, which cannot be fitted by Zeeman profiles. Extreme components in R Leo cannot be reproduced either. 

We may speculate that $\mu$ Cep, U Her, and R Aql exhibit less $V$  components because of weaker SiO maser emission, compared to that of other objects, i.e., the S$/$N is not sufficient to reveal other possible $V$ components. 

\begin{table}
\centering
\caption{Characteristics of Gauss-fitted components to the observed Stokes $V$ spectrum for each source}
\label{tab:V}
\begin{tabular}{ccccccc}
    \hline
    \hline
    Star & \# & v & $\Delta \varv$ & |V$_\textrm{peak}$| & $\delta V$ & S$/$N \\
    && (\kms) & (\kms) & (K) & (mK) &\\
    \hline 
    R Aql & 1 & 49.5 & 2.1 & 0.03 &10& 2.5 \\
    $\mu$ Cep & 1&26.9 & 1.4 & 0.013 &11& 0.9 \\
    &2& 33.5 & 1.9 & 0.018 && 1.2 \\
    $o$ Ceti &1 & 45.3 & 0.9 & 0.13 &29& 13 \\
    &2& 46.9 & 0.4 & 0.22 && 25 \\
    &3& 47.3 & 0.9 &  0.30&& 32 \\
    &4& 48.1 & 0.8 & 0.17 && 18 \\
    &5& 49.0 & 0.2 & 0.10 && 15 \\
     $\chi$ Cyg&1 & 7.3 & 0.9 & 0.11 &34& 10 \\
    &2& 8.6 & 0.5 & 0.31 && 31 \\
    &3& 9.6 & 0.9 & 0.42 && 40 \\
    &4& 10.7 & 0.9 & 0.13 && 13 \\
    &5& 11.9 & 0.7 & 0.29 && 29 \\
    &6& 12.8 & 0.9 & 0.08 && 8.0 \\
    U Her &1& -18.0 & 1.3 & 0.024 &11& 2.0  \\
    (March) &2& -16.1 & 0.7 & 0.04 && 3.5  \\
    &3& -14.7 & 0.7 & 0.026 && 2.1  \\
    U Her &1& -16.9 & 1.3 & 0.11 &25& 2.5 \\
    (May) &  &  &  &&  &  \\
    R Leo &1& -3.5 & 0.9 & 0.21 &29& 33 \\
    &2& -2.3 & 0.9 & 0.19 && 30 \\
    &3& -0.9 & 0.9 & 0.33 && 51 \\
    &4& -0.7 & 0.4 & 0.24 && 42 \\
    &5& 1.5 & 0.9 & 0.44 && 68 \\
    &6& 2.3 & 0.6 & 0.17 && 26 \\
    \hline
\end{tabular}
\tablefoot{\# is the number associated to each component. v is the central position of the gaussian function, $\Delta v$ is the separation between the two peaks of the Zeeman pattern (see Eq. \ref{eq:2sol}) (later called "a" and "b"), |V$_\textrm{peak}$| is the intensity of the peak, $\delta V$ the noise associated to the component, calculated in Sect. \ref{sec:uncert}, and S/N the signal-to-noise ratio (relative to the integrated area).}
\end{table}

\subsection{Maser line saturation}
\label{sec:saturation}

The cosmic maser line theory predicts an exponential growth of the line intensity until the population inversion reaches a critical saturation level, beyond which the radiated emission increases linearly along the propagation axis and tends to be ‘beamed' \citep[see e.g.][]{Goldreich74}. A convenient way to estimate if the maser saturation conditions apply consists in comparing the stimulated emission rate of the maser transition, $R$, with the decay rate, $\Gamma$, from the maser levels. Saturation is observed when R exceeds $\Gamma$.

The stimulated emission rate is derived from:
\begin{equation}
    R =\frac{AkT_b\Delta \Omega}{4 \pi h \nu}
\end{equation}

where $k$ and $h$ are the Boltzmann's and Planck's constants, $T_b$ is the brightness temperature, $A = 2.891 \times 10^{-5} s^{-1}$, the Einstein coefficient for spontaneous emission of transition SiO v= 1, J=2-1, $\nu$ the line frequency, and $\Delta \Omega$ the beaming solid angle (the ratio of the surface area of a typical maser spot to the area of the masing cloud).

Deriving $R$ for the v= 1, J=2-1 SiO transition depends critically on the estimate of $\Delta \Omega$ and $T_b$. $\Delta \Omega$ was estimated from the interferometric maps of SiO at 43 GHz in the v=1, J=1-0 transition observed in one typical Mira star over several epochs \citep{Assaf}; the average over all epochs is about 5 $\times 10^{-2}$, but we note that the minimum and maximum beaming can vary from 2 $\times 10^{-2}$ to nearly 0.1 at different epochs. We assume here that $\Delta \Omega = 5 \times 10^{-2}$ is also representative of the beaming angle for the SiO v=1, J=2-1 transition. To estimate $T_b$ in the absence of high spatial resolution maps of the observed sources, we can rely on the existing measurements made in a broad variety of late type stars. Adopting here 1\,au for the maser typical sizes as in \citet{Perez}, $T_b$ then depends on the distance to the source and the observed total flux density. The estimated $T_b$ is of the order of 1-160 $10^7$ K (see Table \ref{tab:I}, with two different values for two possible distances in R Aql and $\mu$ Cep), 
strongly concurring with the maser nature of these emissions. Derived values of $R$ are given in  Table \ref{tab:I}.

The decay rate from the maser levels may be grossly estimated without describing any pumping scheme, provided we recognize that collisional excitation of vibrational states followed by radiative decay is essential to explain the observed SiO masers in the v=1 state and above \citep{Elitzur92}. At 86 GHz in the v=1 state, we thus get an approximate value of $\Gamma$ from the spontaneous emission rate A(v=1$\rightarrow$0) = 5 s$^{-1}$ \citep{Hedelund} which remains much higher than the collisional de-excitation rate, C(v=1-0)~ a few times $10^{-3}$ to $10^{-4}$ s$^{-1}$ for temperature and density conditions expected in regions where SiO is observed. Of course, the density must remain below very roughly $10^{12}$ cm$^{-3}$ to avoid collisional quenching of the SiO maser.

Despite the uncertainties on the values of $R$ and $\Gamma$, we thus get an estimate of the degree of saturation of the emitting maser for each intensity component and each star (see Table \ref{tab:I}). We hence propose which stars are in the unsaturated ($R < \Gamma$) or in the saturated cases ($R > \Gamma$). The result is used to prefer either one or other of the theories described in Sect. \ref{sec:mg} to compute the magnetic field.

Our results show that only $\chi$ Cyg, $\mu$ Cep (in the case of the larger distance), and U Her are saturated. But considering that we derived lower limits, we deem those objects to be strongly saturated and the others to be just saturated. It is commonly admitted in the literature that circumstellar SiO masers are in the saturated or strongly saturated regime \citep[e.g.,][]{Kemball2009}. We will nevertheless consider both cases for each star.

\subsection{Magnetic field estimate}
\label{sec:mg}

The computation of the magnetic field strength via the Zeeman effect differs according to the saturation of the maser emission. The magnetic field is computed from the two peaks found for Stokes $V$ with our \textit{rvm} code. With the Gaussian decompositions used in Eq. \ref{eq:I} the intensity extrema are observed at:
\begin{equation}
    \varv_{peak} = \varv \pm \frac{1}{\sqrt{2}} \Delta \varv_\mathrm{D} \label{eq:2sol}
\end{equation}
the two peaks, noted a and b (see Table \ref{tab:VE}), being separated by $\Delta \varv = \sqrt{2} \Delta \varv_\mathrm{D}$ (see Table \ref{tab:V}).

\subsubsection{The case of a saturated maser} 

In a saturated maser, the fractional Stokes vectors propagating through the masing cloud are constant. The radiative transfer equation for the Stokes vectors admits very few situations with a non-zero constant and propagating Stokes vector. These solutions have been thoroughly studied by \citet{Elitzur1996}, while other theories have been proposed \citep[see][for a brief discussion]{Herpin2006}.

Let us write the ratio $x_B$ of the Zeeman shift $\Delta v_B$ to the Doppler line-width $\Delta v_\mathrm{D}$ both in \kms, 
\begin{align}
    x_B &= \Delta \varv_B / \Delta \varv_\mathrm{D} \label{eq:xB}
\end{align}
for the Zeeman effect by explicitly writing $\Delta \varv_B = 14g\lambda_0 B$, where $g$ is the Land\'e factor in terms of the Bohr's magneton (where g, for the diamagnetic molecule SiO, is $8.44\cdot 10^{-5}$, see Appendix \ref{app:lande}), $\lambda_0 = 0.34785$~cm the wavelength of the SiO v=1 J=2-1 maser emission, and $B$ the magnetic field strength in Gauss.

\begin{table*}
\centering
\caption{Polarization rates, angles, and magnetic field values for all components and all stars for both saturated and unsaturated cases.} 
\label{tab:VE}
\begin{tabular}{crccccccc}
\hline
\hline
    Star & \# & $p_\mathrm{L}$ & $p_\mathrm{C}$ & $\chi$ & $B\textsubscript{E}$ & $B\textsubscript{E}$ (2$^\textrm{nd}$ solution)  & $B\textsubscript{Z} \times \textrm{cos}({\theta}_2)$ & Comment \\
    & & (\%) & (\%) & ($^\circ$) &(Gauss) & (Gauss) & (Gauss)& \\
    \hline
R Aql &1 - a & 7.5 & -1.0 & 179.5 & (6.1 $\pm$ 2.0) & (3.7 $\pm$ 1.3) & (5.3 $\pm$ 1.8)&* \\
&  - b & 21.4 & 3.4 & 95.3 & (21.6 $\pm$ 7.3) & (9.8 $\pm$ 3.3) & (8.2 $\pm$ 2.8)&*\\
$\mu$ Cep &1 - a & 4.5 & -0.9 & 143.6 & (3.7 $\pm$ 3.4) & (2.4 $\pm$ 2.2) & (12.2 $\pm$ 11.1)&* \\
&  - b & 5.3 & 1.0 & 133.8 & (4.3 $\pm$ 3.9) & (2.7 $\pm$ 2.4) & (28.1 $\pm$ 25.6)&* \\
&2 - a & 7.9 & 14.0 & 145.4 & (76.9 $\pm$ 49.4) & (43.3 $\pm$ 27.8) & (63.6 $\pm$ 40.1)&*\\
&  - b & 8.1 & -1.8 & 117.6 & (10.4 $\pm$ 6.7) & (6.1 $\pm$ 3.9) & (9.9 $\pm$ 6.4)&*\\
$o$ Ceti &1 - a & 58.9 & -2.2 & 147.5 & (8.0 $\pm$2.1)  & & (40.0 $\pm$ 10.8)&* \\
& - b & 9.6 & 6.2 & 136.4 & (23.4 $\pm$ 6.3) & (13.2 $\pm$ 3.6) & (182.5 $\pm$ 49.5)&*\\
&2 - a & 33.4 & 3.3 & 76.1 & 2.5 $\pm$ 0.4 && 2.7 $\pm$ 0.42 &\\
&  - b & 15.9 & 0.8 & 77.1 & 1.6 $\pm$ 0.3 &0.8 $\pm$ 0.1& 5.2 $\pm$ 0.8&\\
&3 - a & 23.3 & 2.1 & 65.6 & 4.9 $\pm$ 0.5 &2.1 $\pm$ 0.2& 31.0 $\pm$ 3.4 &\\
&  - b & 4.7 & -1.6 & 29.9 & 8.9 $\pm$ 1.0 &5.8 $\pm$ 0.6& 13.2 $\pm$ 1.5&\\
&4 - a & 4.7 & -1.6 & 29.9 & 4.2 $\pm$ 0.9 &2.7 $\pm$ 0.8& 7.0 $\pm$ 1.5 &\\
&  - b & 29.6 & -2.2 & 33.9 & 4.0 $\pm$ 0.8 &1.5 $\pm$ 0.3& 5.7 $\pm$ 1.2&\\
&5 - a & 47.0 & -1.5 & 35.3 & 0.4 $\pm$ 0.2 && 1.0 $\pm$ 0.4 &\\
&  - b & 52.6 & 0.0 & 36.4 & 0.3 $\pm$ 0.1 && 0.8 $\pm$ 0.3&\\
$\chi$ Cyg &1 - a & 26.2 & -1.7 & 38.7 & 4.3 $\pm$ 1.5 &1.7 $\pm$ 0.6& 10.1 $\pm$ 3.6 &\\
&  - b & 16.2 & 1.1 & 12.4 & 3.0 $\pm$ 1.1 &1.5 $\pm$ 0.5& 7.5 $\pm$ 2.7&\\
&2 - a & 36.4 & 1.9 & 26.7 & 2.6 $\pm$ 0.3 && 2.4 $\pm$ 0.3 &\\
&  - b & 50.2 & 0.0 & 33.9 & 0.9 $\pm$ 0.1 && 1.2 $\pm$ 0.1&\\
&3 - a & 51.8 & 0.2 & 33.9 & 1.6 $\pm$ 0.2 && 2.5 $\pm$ 0.2 &\\
&  - b & 39.4 & -0.8 & 34.8 & 3.3 $\pm$ 0.3 && 2.6 $\pm$ 0.2&\\
&4 - a & 32.8 & -1.0 & 32.8 & 1.4 $\pm$ 0.4 &1.4 $\pm$ 0.4& 1.8 $\pm$ 0.5 &\\
&  - b & 26.2 & -1.2 & 25.9 & 1.8 $\pm$ 0.5 &0.7 $\pm$ 0.2& 2.4 $\pm$ 0.7&\\
&5 - a & 38.9 & -1.2 & 34.0 & 2.1 $\pm$ 0.3 && 5.3 $\pm$ 0.7 &\\
&  - b & 30.4 & 1.7 & 30.2 & 4.3 $\pm$ 0.6 &1.6 $\pm$ 0.2& 2.6 $\pm$ 0.3&\\
&6 - a & 28.5 & 2.2 & 28.9 & (2.3 $\pm$ 1.1) & (0.9 $\pm$ 0.4) & (1.0 $\pm$ 0.5)&* \\
& - b  & 28.6 & 2.2 & 22.3 & (7.7 $\pm$ 3.5) & (3.0 $\pm$ 1.4) & (47.8 $\pm$ 22.1)&*\\
U Her &1 - a & 18.8 & 0.9 & 136.7 & 3.9 $\pm$ 1.8 &1.9 $\pm$ 0.9& 2.2 $\pm$ 1.0 &\\
(March)&  - b & 1.7 & -0.4 & 93.1 & 1.6 $\pm$ 0.7 &1.1 $\pm$ 0.5& 11.8 $\pm$ 5.5&\\
&2 - a & 9.4 & -0.8 & 177.9 & 1.5 $\pm$ 0.4 &0.9 $\pm$ 0.2& 10.5 $\pm$ 2.8 &\\
&  - b & 5.0 & 0.8 & 170.6 & 1.5 $\pm$ 0.4 &1.0 $\pm$ 0.3& 5.9 $\pm$ 1.6&\\
&3 - a & 9.3 & 1.0 & 115.8 & 1.7 $\pm$ 0.8 &1.0 $\pm$ 0.5& 2.3 $\pm$ 1.1 &\\
&  - b & 4.2 & -1.3 & 51.0 & 3.2 $\pm$ 1.5 &2.0 $\pm$ 1.0& 9.8 $\pm$ 4.5&\\
U Her &1 - a & 13.1 & -1.2 & 162.3 & 4.4 $\pm$ 1.0 &2.4 $\pm$ 0.6& 4.9 $\pm$ 1.2 &\\
(May)&  - b & 12.0 & 1.3 & 7.8 & 5.1 $\pm$ 1.2 &2.8 $\pm$ 0.7& 17.1 $\pm$ 4.0&\\
R Leo &1 - a & 31.2 & 2.3 & 79.2 & 6.3 $\pm$ 1.0 &2.4 $\pm$ 0.4& 6.9 $\pm$ 1.1 &\\
&  - b & 32.2 & -1.6 & 21.1 & 8.5 $\pm$ 1.3 &3.2 $\pm$ 0.5& 32.1 $\pm$ 5.0&\\
&2 - a & 10.0 & 0.7 & 15.1 & 5.7 $\pm$ 1.0 &3.3 $\pm$ 0.6& 8.8 $\pm$ 1.5 &\\
&  - b & 8.3 & -1.2 & 152.7 & 2.1 $\pm$ 0.4 &1.2 $\pm$ 0.2& 2.0 $\pm$ 0.4&\\
&3 - a & 7.2 & -0.9 & 164.7 & 1.8 $\pm$ 0.2 &1.1 $\pm$ 0.1& 3.9 $\pm$ 0.4 &\\
&  - b & 13.4 & 1.2 & 72.2 & 2.0 $\pm$ 0.2 &1.1 $\pm$ 0.1& 8.1 $\pm$ 0.8&\\
&4 - a & 5.5 & -0.7 & 179.7 & 0.4 $\pm$ 0.1 &0.3 $\pm$ 0.1& 7.6 $\pm$ 1.2 &\\
&  - b & 13.4 & 1.2 & 72.2 & 0.6 $\pm$ 0.1 &0.3 $\pm$ 0.1& 5.2 $\pm$ 0.8&\\
&5 - a & 19.1 & 1.1 & 164.1 & 2.9 $\pm$ 0.2 &1.4 $\pm$ 0.1& 5.1 $\pm$ 0.4 &\\
&  - b & 21.2 & -1.6 & 166.2 & 6.3 $\pm$ 0.5 &2.8 $\pm$ 0.2& 11.2 $\pm$ 0.9&\\
&6 - a & 16.9 & -1.8 & 171.0 & 1.7 $\pm$ 0.4 &0.9 $\pm$ 0.2& 6.2 $\pm$ 1.3 &\\
&  - b & 16.5 & -1.6 & 173.9 & 1.8 $\pm$ 0.4 &0.9 $\pm$ 0.2& 13.6 $\pm$ 2.8&\\
\hline
\end{tabular}
\tablefoot{\# is the number of the components described in Table \ref{tab:V}, the two peaks for each component are represented by "a" and "b", separated by $\Delta v$. $p_L$, $p_C$, and $\chi$ are respectively the linear polarization, the circular polarization and the angle of polarization. $B_E$ is the magnetic field value computed with Eq. \ref{eq:Elitzur}, the second solution for $p_L < 33\%$ (see Sect. \ref{sec:angle}). $B_Z$ in column 8 is resulting from Eq. \ref{eq:Zeeman} and is still depending on the angle. The $\ast$ in the last column indicates unreliable results.}
\end{table*}

Under the saturation regime and for the ratio $x_B<1$ (which is appropriate for diamagnetic molecular species), \citet{Elitzur1996} found that $x_B$ is proportional to the fractional circular polarization $p_\textrm{C}$ at the peak of the Stokes $V$ parameter according to the law:
\begin{align}
    x_B &= \frac{3\sqrt{2}}{16} p_\textrm{C} \cos \theta_1
    \label{eq:xB2}
\end{align}
where $\theta_1$ is the angle between the magnetic field direction and the incident beam arriving at the SiO molecule. 

\citet{Herpin2006} have followed a similar approach to estimate the magnetic field strength from their observed values of $p_\textrm{C}$. However, they used $g=10^{-3}$, so that their field strength estimates must be reevaluated using $g=8.44\cdot 10^{-5}$ for comparison with our new study (see Sect. \ref{sec:disc}). From $\Delta \varv_B$ above and Eq. \ref{eq:xB2} we get at the peak, $\varv_\mathrm{peak}$, of the Stokes $V$ parameter:
\begin{align}
    \frac{B}{\cos \theta_1} &\simeq 645 \Delta \varv_\mathrm{D} p_\textrm{C}(\varv_\mathrm{peak}) \label{eq:Elitzur}
\end{align}
Eq. \ref{eq:Elitzur} is used to derive the magnetic field strength in Gauss, noted $B\textsubscript{E}$ in the following, when $p_\textrm{C}$ and $\Delta \varv_\mathrm{D}$ are known. The remaining uncertainty in the determination of the field intensity comes from $\theta_1$ discussed in Sect. \ref{sec:angle}.

\subsubsection{The case of an unsaturated maser}

Whenever the maser is unsaturated and the fractional Stokes vectors are not constant along the optical path, the typical weak-field approximation applies. The Zeeman frequency shift of a given level is proportional to the Larmor frequency,
\begin{align}
    \nu_L = \frac{\mu}{h} B ,
\end{align}
where $h$ is the Planck's constant. The magneton $\mu$ is often the Bohr's magneton. But for the diamagnetic molecule SiO all electronic angular momenta are zero and the only contribution to the total angular momenta comes from angular momenta associated with the rotation of the total mass of the molecule and those momenta are expressed in units of the nuclear magneton. It is however not uncommon to rewrite formulae in terms of the Bohr magneton and re-define accordingly the Land\'e factor. Details can be found in the Appendix \ref{app:lande}. In what follows we adopt such a convention and, as in the case of the saturated maser, we write all formulae in terms of the Bohr magneton.



Under the weak field approximation, it can be demonstrated that the expected Stokes $V$ line profile is proportional to the derivative of the intensity profile (see also Eq. \ref{eq:Vrvm}): 
\begin{align}
    V(\varv) &= \Delta \varv_B \textrm{cos}\theta_2 \frac{d I}{d \varv} (\varv)
\end{align}
where $\theta_2$ is the angle between the magnetic field direction and the emitted beam (towards the observer). We can express $V$ as a function of $B$ with the velocity axis sampled with 0.139 km$/$s channel separation.
\begin{equation}
    V(\varv) \simeq 14 g \lambda_0 B \cos \theta_2 \frac{d I}{d \varv} (\varv)  \frac{1}{0.139} 
\end{equation}




Once again, we assume that the line profile of each maser component is Gaussian, allowing us to retrieve the magnetic field strength from the peak amplitudes of Stokes $V$ and $I$:
\begin{align}
    B \cos  \theta_2 &\simeq 338 \frac{V(\varv_\mathrm{peak})} {d_{\varv}I(\varv_\mathrm{peak})}  \label{eq:Zeeman}
\end{align}

$d_{\varv}$ being the derivative with respect to the velocity $= \frac{\delta}{\delta_{\varv}}$.\\

In the unsaturated case, the magnetic field strength will be noted $B\textsubscript{Z}$ in the following.

\begin{figure}
    \centering
    \includegraphics[width=0.95\columnwidth]{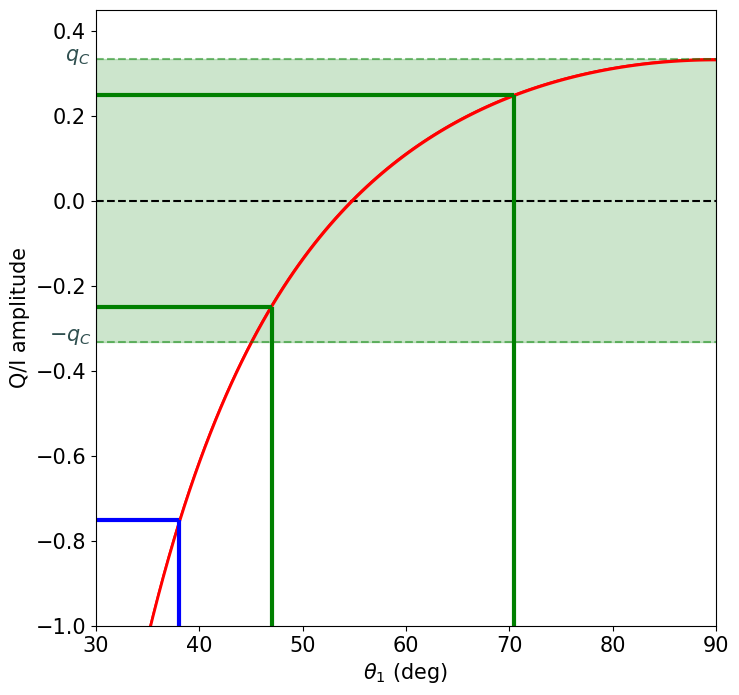}
    \caption{Ratio $q=Q/I$ vs the angle $\theta_1$ (Eq. \ref{eq:Q}). In the green zone ($|q| < q_c$), there are two solutions for the angle, as we can see with the example with $|q|=0.25$, the two green lines. Out of this zone (example in blue) there is one solution only.}
    \label{fig:linearpolar}
\end{figure}

\subsubsection{Determination of the angle of incidence of the maser beam}
\label{sec:angle}

The two different angles, $\theta_1$ and $\theta_2$, defined earlier for the saturated and unsaturated cases respectively are not easily determined unless we have access to independent information on the field direction. As $x_B$ is less than one, all Zeeman components overlap and only the 'Type 0' polarization case described in \citet{Elitzur1996} applies. Following Elitzur's reasoning, in the saturated case, we can compute angle $\theta_1$ using our linear polarization results.

Under the weak field approximation, there is no linear polarization in the maser line \citep[see review by][]{Watson2009}. A second order approximation can be proposed for linear polarization, but it does not describe correctly the linear polarization profiles. Linear polarization due to the Zeeman effect is in general one order of magnitude smaller than circular polarization \citep[][]{Perez} and has zero net polarization over wavelength \citep[see][]{Landi2004}. One could claim that we observe the very specific geometries that afford large amplitudes to linear polarization under the Zeeman effect: fields transverse to the line of sight. But accepting that we systematically observed transverse fields appears as incongruous. And in no case the Zeeman effect under the weak field approximation can produce Gaussian-like profiles for linear polarization as those observed here \citep[e.g.][]{Wiebe1998, Watson2009}. Faraday rotation \footnote{We mean in-source Faraday rotation resulting in a modification of the radio polarization properties in the presence of a magnetic field. Electrons mixed up with the SiO cloud and in the circumstellar envelope could rotate the SiO linearly polarized wave; the latter effect is hopefully small given the short distances involved, unless the field intensity is large.} can produce net linear polarization, but in order to do so the magnetic fields involved would be too large to be described under the weak field approximation, and then it is circular polarization that could not be explained neither in its amplitude nor in its shape. 
We can safely conclude that, under the hypothesis of the weak field approximation, the observed linear polarization must be due to another physical mechanism than the Zeeman effect, as we further discuss in section \ref{sec:Icomp}. 
On the other hand, Elitzur's theory shows that a linear polarization signal is to be expected together with the circular polarization one. Once more assuming a Gaussian line profile, and following \citep{Elitzur1996}, in a coordinate system that is aligned with the sky-plane projected magnetic field, $q = Q/I$ is given by: 
\begin{equation}
     q = 1 - 2\frac{3- 2R_1}{(2R_1 + 1)\textrm{sin}^2\theta_1 } \label{eq:Q}
\end{equation}
with $R_1$ a function of $x_B$ (defined in Eq. \ref{eq:xB}) and of the dimensionless frequency argument $x$ defined in \citet{Elitzur1996}. The numerical values of $x_b$ being small, $R_1$ is not sensitive to the exact values of $\Delta \varv_B$ and $\Delta \varv_D$ and this is used below to derive the critical value of $q$ when $\theta_1$ approaches $90^\circ$ near the SiO line center where $x=0$.
\begin{equation}
    R_1 = 1 - x_B^2(1-2x^2)
\end{equation}

The dependence of $q$ with $\theta_1$ is shown in Fig \ref{fig:linearpolar}. Given our typical S$/$N ratios, and due to the quadratic dependence of $B$ and the weak values of the field strength expected, the relationship shown on this figure is independent of field strength. A value of $q$ provides us with an independent estimate of $\theta_1$.

This linear polarization $q$ is perpendicular to the plane formed by the incident beam direction and the magnetic field vector, and no other linear polarization signal is expected. We ignore what this direction is in any maser component, and we therefore ignore how this linear polarization is projected into our instruments directions for Stokes $Q$ and Stokes $U$. But the total polarization $p_\mathrm{L}$ measured must be the total amplitude of the emitted linear polarization $q$ in Elitzur's theory, except for its sign.
Since we only have access to the absolute value of $p_\mathrm{L}$ from our observations, two different situations may occur depending on the measured value of $p_\mathrm{L}$ with respect to a critical value $q_C$ defined by $\lim\limits_{\theta_1 \rightarrow +90^\circ} q = 0.33$:
\begin{itemize}
    \item $|q| > q_c$: the solution for the angle  $\theta_1$ is unique (see example in blue in Fig. \ref{fig:linearpolar}).
    \item $|q| < q_c$, there are two solutions for $\theta_1$ (see example in green), and hence  two solutions for the magnetic field strength.
\end{itemize}

\section{Results}
\label{sec:results}

\subsection{Observed polarization and maser excitation}
\label{sec:Icomp}

The circular and linear polarization fractions derived for each identified $V$ component in each star are given in columns 6 and 7 of Table \ref{tab:VE} for each detected component. The other columns present the magnetic field estimates that will be discussed in the next section.

Two regimes for the linear polarization are observed in our sample. $\mu$ Cep, R Aql, and U Her mostly exhibit low levels of the  $p_\mathrm{L}$ parameter ($<25$\%) while other stars show high $p_\mathrm{L}$ values, up to 60\%. Our polarization levels can be compared with those of \citet{Herpin2006}. The linear polarization fractions are similar in these two works.  However, while they measured an average absolute value of circular polarization of 9\% and 5\% in Miras and semi-regular variables, respectively, we have only 1.9\% in our sample (composed of 5 Miras and the RSG $\mu$ Cep). We also note that several components in Herpin et al. have $p_\mathrm{C}$-values around or below 2\%. The lower level of polarization observed in supergiants \citep{Herpin2006} is also confirmed in $p_\mathrm{L}$ for $\mu$ Cep but not in circular polarization as this star exhibits the highest $p_\mathrm{C}$ fraction in our sample. A specific comparison can be done for R Leo and $\chi$ Cyg observed in both studies, even if the Stokes $I$ maser line profile has deeply changed between the two observation epochs. The linear polarization fraction reaches the same 30\% and 40\% level for R Leo and $\chi$ Cyg (with some higher values for some components). The circular polarization is definitely lower in our case (2.3\% maximum, compared to 10-20\%).  

These values might shed light on the origin of the observed polarization in relation with the Zeeman or other effects. We can investigate which mechanisms could produce the observed linear polarization fractions. 
In the case of isotropic pumping of the energy levels giving rise to maser emission \citep[e.g.][]{Nedoluha1990} and for a moderately saturated maser, $p_\mathrm{L}$ greater than 15\%, up to 25\%, can be generated only with peculiar angle configurations (angle ${\theta}_2$ larger than $45^{\circ}$). Under anisotropic pumping conditions, much higher $p_\mathrm{L}$ values can be achieved, but no high $p_\mathrm{C}$ values can be generated \citep[see][]{Lankhaar2019}. Actually, non-Zeeman mechanisms may also lead to circular polarization. \citet{Asensio2005}, considering the Hanle effect, showed that changes in the radiation anisotropy conditions can result in the rotation of the electric vector position angle (EVPA) and in the anisotropic pumping of a ‘dichroic’ maser. Several variations of anisotropic pumping mechanisms \citep[e.g.][]{Western1983,Wiebe1998, Watson2009} or anisotropic resonant scattering of a magnetized foreground SiO gas \citep{Houde2014}, have been proposed for the non-Zeeman case too. Such non-Zeeman mechanisms have been ruled out in some specific environments, and on the basis of very high spatial resolution observations (VLBI) of the EVPA \citep[see e.g.][]{Kemball2009,Richter2016,Tobin2019}. Several polarized maser theories have successfully been developed but,  obviously, none can fully explain all the cases observed.

\begin{figure}
    \centering
    \includegraphics[width=1.0\columnwidth]{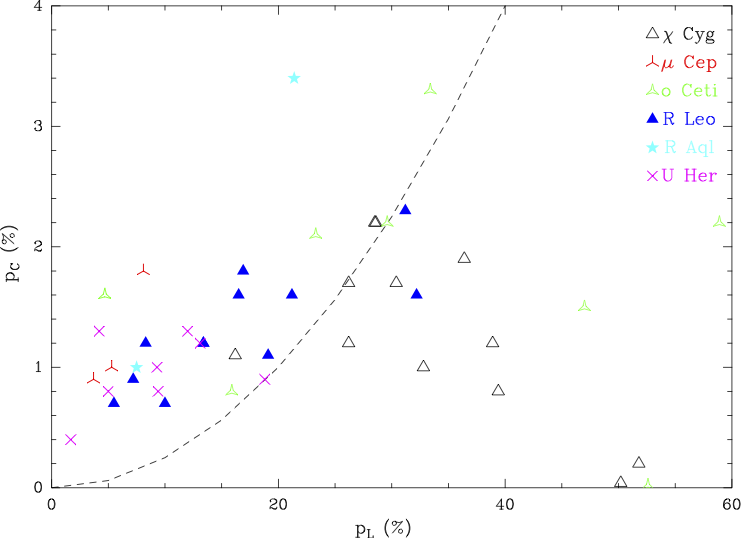}
    \caption{Absolute fractional polarization $p_\mathrm{C}$ versus $p_\mathrm{L}$ for the whole sample and all maser components. The dashed line corresponds to $p_\mathrm{C} = {p_\mathrm{L}}^2/4$.}
    \label{fig:pc_pl}
\end{figure}

As indicated by \citet{Wiebe1998}, non-Zeeman mechanisms in unsaturated masers should lead to a proportional relation between $p_\mathrm{C}$ and $p_\mathrm{L}$  (as $p_\mathrm{C}$ would be created from $p_\mathrm{L}$) with $\overline{p_\mathrm{C}} \leq {\overline{p_\mathrm{L}}}^2/4$. \cite{Herpin2006} found a linear function between the two polarizations, but with a large spread. Actually, our sample does not reveal any correlation (see Fig. \ref{fig:pc_pl}, except maybe for R Leo).  The second condition ($\overline{p_\mathrm{C}} \leq {\overline{p_\mathrm{L}}}^2/4$) is not fulfilled either (see Table \ref{tab:pmoyen}), as also observed by \citet{Richter2016} and \citet{Tobin2019}, except for $\chi$ Cyg. 
Considering that the maser emission in this star is definitely saturated (see Table \ref{tab:I}), we adopt the Zeeman scenario for the whole sample. Nevertheless, we note that the presence of a velocity gradient along the propagation path, but still consistent with coherent  SiO maser amplification in some sources, could enhance $p_\mathrm{C}$ and might even produce non anti-symmetric Stokes $V$ spectra \citep{Nedoluha1994}.

Now we consider the polarization angle $\chi$ as defined by Eq. \ref{eq:param}. First, we underline that owing to our limited spatial resolution, the underlying Stokes $Q$ and $U$ parameters are beam-averaged quantities. Hence, if there were two spatially separated maser spots within our main beam, each one with a different EVPA but emitting at the same velocity, we would not be able to make any rigorous interpretation of the measured $\chi$ angle. Only VLBI observations could separate two such maser spots and accurately trace the EVPA throughout the photosphere. 
As explained by \citet{Tobin2019}, an angle between the maser propagation and the line of sight  smaller or larger than the van Vleck angle ($55^{\circ}$) will determine if the magnetic field is, respectively, parallel or perpendicular to the EVPA. As in past studies \citep[e.g.][]{Herpin2006,Kemball2011, Tobin2019}, some $\chi$ jumps are observed for some sources across the SiO profile: R Leo ($\sim70^{\circ}-160^{\circ}$), $o$ Ceti ($\sim35^{\circ}-70^{\circ}-140^{\circ}$), and R Aql ($\sim95^{\circ}-180^{\circ}$). The behavior in U Her is more erratic, i.e., we do not observe beam-averaged $\pi/2$ jumps while clear $\pi/2$ rotations are reported in VLBI publications; again, this can be explained by beam-smearing effects in our data. Such a rotation together with the dependence of the fractional linear polarization with the angle between the magnetic field and the line of sight is predicted in 'linear' and saturated masers that are isotropically pumped \citep{Goldreich1973, Elitzur2002}. These models  and the VLBI polarimetric observations of TX Cam \citep{Kemball2011} tend to rule out, at least in this object, the model proposed by \citet{Asensio2005}. However, we also point out that although the Goldreich et al. (1973) maser model seems adequate, some Faraday rotation must be invoked to explain the smooth rotation of the EVPA  \citep[see][]{Kemball2011, Tobin2019}. Finally, we mention that changes in the EVPA could also result from rapidly changing magnetic field orientations above cool magnetic spots on the stellar photosphere \citep{Soker2002} (assuming that there are physical connections between photospheric spots and the upper atmospheric layers where SiO is observed).

\subsection{Magnetic field strength}
\label{sec:mfc}
\begin{figure*}
    \centering
    \includegraphics[width=0.33\textwidth]{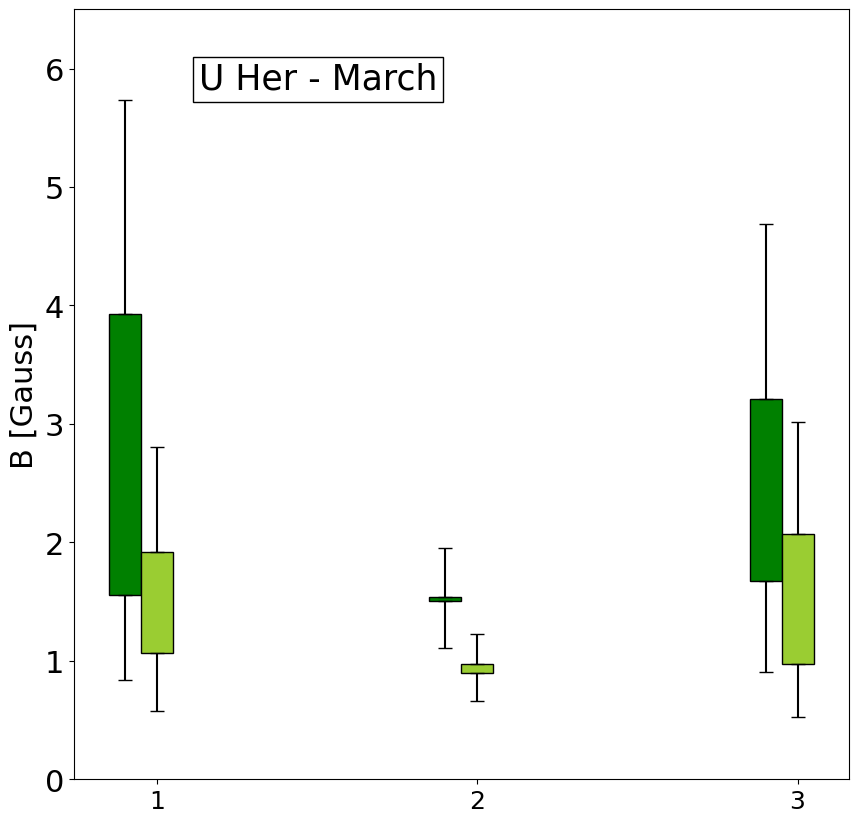}
    \includegraphics[width=0.33\textwidth]{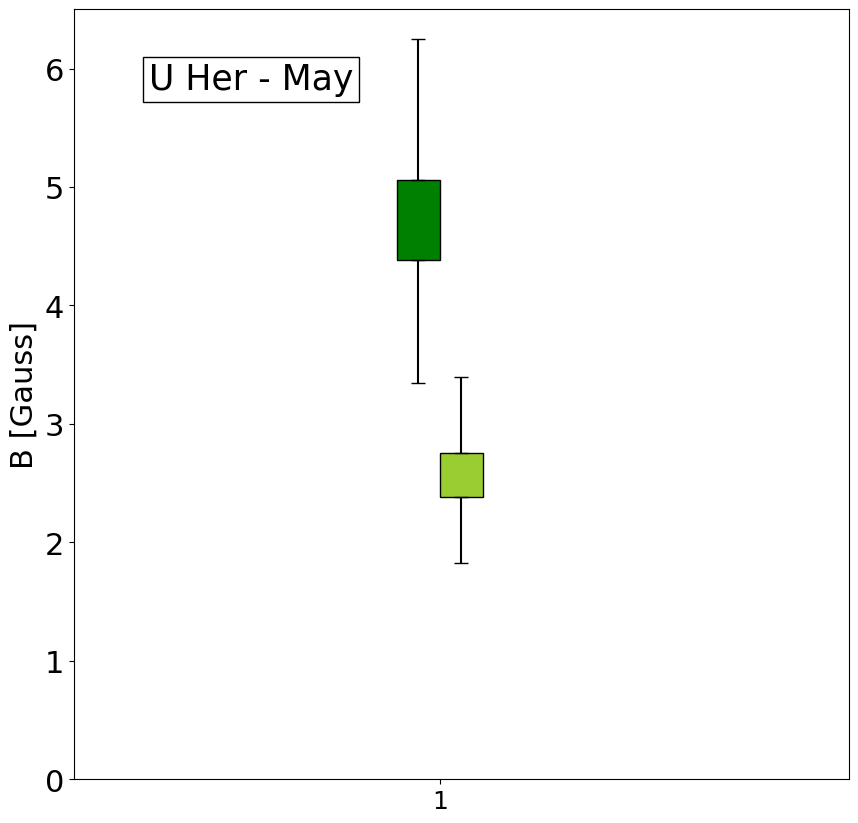}
    \caption{Magnetic field values for U Her for the two observation periods, computed from Eq. \ref{eq:Elitzur} with two $\theta_1$ angle solutions, (dark and light green), when the linear polarization is below $33\%$ (see Sect. \ref{sec:angle} and Fig. \ref{fig:linearpolar}). On the x-axis, the numbers correspond to the Gaussian components, as described in the second column of the Table \ref{tab:VE}. The bar length corresponds to the difference between the magnetic field computed at the first peak of the derivative of the Gaussian component and the second one.}
    \label{fig:Uher}
\end{figure*}

Applying the method described in Sect. \ref{sec:mg} we derive for each star estimates of the line-of-sight magnetic field strength $B\textsubscript{E}$ and $B\textsubscript{Z} \times \textrm{cos}({\theta}_2)$ for both the saturated and unsaturated cases (see Table \ref{tab:VE}, Fig. \ref{fig:Uher}, and Appendix \ref{sec:Appen:MF}). As explained in Sect. \ref{sec:StokesDec}, we compute the magnetic field strength at the two peaks of each detected anti-symmetric $V$ profile, which results in two extreme values of $B$ for each Zeeman component ("a" and "b" in the Table \ref{tab:VE}). Values in the unsaturated case are generally larger. 
 
While the angle ${\theta}_2$ in the unsaturated case remains unknown, in the saturated case the angle ${\theta}_1$ has been estimated, with sometimes two solutions when the linear polarization is lower than the critical value (see Sect. \ref{sec:angle}), hence leading to two possibilities for $B\textsubscript{E}$. These two solutions for the same angle $\theta_1$ have been used to derive $B\textsubscript{E}$ in Table \ref{tab:VE}. We cannot exclude that the observed linear polarization might include a contribution from non-Zeeman mechanisms, hence leading to some overestimation of $p_\mathrm{L}$, generated under the action of the Zeeman effect and hence of ${\theta}_1$; in that case $B\textsubscript{E}$ values in Table \ref{tab:VE} should be considered as lower limits. 

At least two objects, $\chi$ Cyg and U Her can be considered as giving rise to saturated masers (see columns 6 and 7 in Table \ref{tab:VE}). For $\chi$ Cyg, B\textsubscript{E} varies between 0.7 and 4.3 Gauss for components 1-5. We note that the results $B\textsubscript{Z} \times \textrm{cos}({\theta}_2)$ for the unsaturated case are of the same order of magnitude (a bit higher for component \# 1). Results for component \#6 are less reliable, considering the large uncertainty and difference in the estimate between the two peaks a and b (see Figure \ref{fig:Appen_sat}). We then exclude this component in the further discussion.  For U Her, the magnetic field (B\textsubscript{E}) is around 0.9-1.5 Gauss in March and between 2.4-5.1 Gauss in May considering the central component \#2, at a similar velocity for both periods. The two other side components (\#1 and \#3) observed in March have large uncertainties and are not detected in May (see Fig. \ref{fig:Uher}).

Considering that all other objects exhibit unsaturated maser emission (see Sect. \ref{sec:saturation}), we assume that $B\textsubscript{Z} \times \textrm{cos}({\theta}_2)$ provides a reliable estimate of the field although ${\theta}_2$ remains unknown; indicative ad-hoc values of ${\theta}_2$, for instance $45^{\circ}$, can be considered. Estimates of the magnetic field (assuming ${\theta}_2=45^{\circ}$) differ from star to star and component to component. In the case of R Aql, the Stokes $V$ spectrum is largely dominated by the instrumental leakage of Stokes $I$ into Stokes $V$ (see Sect. \ref{sec:instr_I_V}). We therefore consider that $B\textsubscript{Z} \times \textrm{cos}({\theta}_2) = 8.2$ Gauss as given in Table \ref{tab:VE} is an upper limit of the field intensity in the unsaturated case. For the two objects R Leo and $o$ Ceti, the magnetic field in terms of the quantity $B\textsubscript{Z} \times \textrm{cos}({\theta}_2)$ is  2.0-32.1 and 0.8-31.0 Gauss, respectively. The component \#1 of $o$ Ceti has been discarded because we did not succeed in fitting this component with the derivative of a Gaussian function (see Sect. \ref{sec:StokesI} and Fig. \ref{fig:V_fitted}).Therefore, it may not be shaped by any Zeeman effect, in which case the estimated magnetic field would not be relevant.



Because of the very low S$/$N for the $V$ components in $\mu$ Cep (see Table \ref{tab:V}), the field uncertainties are too large and not taken into account in our discussion.




\section{Discussion}
\label{sec:disc}

Since the 1990s, various scenarios implying slow stellar rotation, the presence of a companion, and a weak magnetic field have been proposed to tentatively explain how the geometry of the CSE can be shaped in evolved stars \citep[see for instance ][]{Soker1992}. All studies report a too weak magnetic field strength to dominate the dynamics of the gas \citep[][]{Balick2002, Herpin2006, Vlemmings2017}. Nevertheless, a moderate field of the order of 1 Gauss at the stellar surface can play the role of a catalyst or a collimating agent \citep[][]{Soker1998, Blackman2001, Greaves2002}. Actually the role of a companion in shaping the envelope, without the need to invoke a magnetic field, has been recently demonstrated by \citet{Decin2020}. 

The strength of the magnetic field decreases with increasing distance $r$ from the photosphere and can be expressed as a power law that depends on its geometry \citep[][]{Parker1958, Matt2000, Meyer2021}, in $r^{-1}$, $r^{-2}$, and $r^{-3}$ for a toroidal, poloidal, and dipole field respectively. When comparing their results to field estimates derived from the OH and H$_2$O masers farther away in the CSE, \citet{Herpin2006}, \citet{Duthu2017}, and \citet{Vlemmings2017} concluded that a toroidal magnetic field was dominant (see also later in this section). A combination of a toroidal and a poloidal field is nevertheless not excluded \citep[see ][for instance for RSGs]{Meyer2021}.

\begin{figure}
    \centering
    \includegraphics[width=0.45\textwidth]{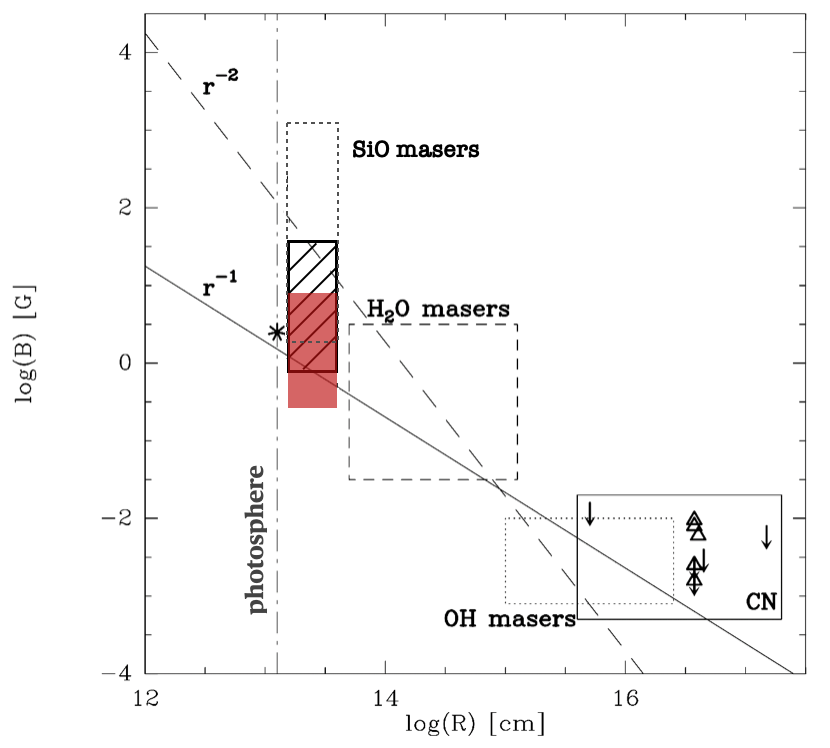}
    \caption{Evolution of the field strength with the distance \citep[adapted from][]{Vlemmings2012,Duthu2017}. The four boxes show the SiO masers (the red box represents the results of this work for saturated maser, the hatched box for unsaturated maser using $\theta_2=45^{\circ}$, the dotted box the results from \citet{Herpin2006}), the H$_2$O and OH masers, and CN. The $r^{-2}$ solar-type and $r^{-1}$ toroidal magnetic fields are discussed in Sect. \ref{sec:disc}. The star close to the SiO maser box is for the magnetic field measured by \citet{Lebre2014} in $\chi$ Cyg. The vertical dashed line indicates the stellar surface for the Miras.}
    \label{fig:Bevol}
\end{figure}

In the following, we compare our newer observations and calculations to those in  \citet{Herpin2006}, who had considered saturated maser emission only. First of all, we point out that the field intensities derived in Herpin et al. must be increased because an erroneous Land\'e factor ($g = 10^{-3}$ instead of $8.44 \times 10^{-5}$) and an erroneous wavelength were used; the corrective factor  is 9.924. The magnetic field strengths derived here are thus lower than in Herpin et al. (2006) for the same circumstellar region where the SiO radio emission is observed (2-5 R$_{\star}$ from the photosphere according to SiO VLBI map). \citet{Herpin2006} found a mean value of $\sim$34 Gauss (which includes the corrective factor above), while our current estimates are below 32 Gauss (excluding the unreliable results in Table \ref{tab:VE}), and show a mean value of 2.5 Gauss for the saturated case, and 12.4 Gauss for the unsaturated one, assuming $\theta_2 = 45^{\circ}$. If we consider, as in \citet{Herpin2006} the saturated case only, our magnetic field estimates range from 0.3 to 8.9 Gauss. We now focus the comparison on $\chi$ Cyg and R Leo, observed in both studies. While for R Leo (adopting $\theta_1=45^{\circ}$) B was estimated by \citet{Herpin2006} to be, after re-evaluation, 41-45 Gauss in the saturated case, we find somewhat lower values here: $B\textsubscript{E} \sim 0.3 - 8.5$ G and $B\textsubscript{Z} \sim 2.8 - 4$ G (at $\theta_2 = 45^{\circ}$ for the 2.0-32.1 G interval in Table \ref{tab:VE}) for the saturated and unsaturated cases, respectively. Of course, and this argument stands for all objects, both observations are separated by more than twenty years and not only the stellar activity but also the magnetic field may have changed during this period. We nevertheless stress that, despite uncertainties in the degree of maser saturation (resulting in an uncertainty to use on the B-formula), our careful calibration method and Zeeman component fitting procedure lead to reasonable magnetic field estimates. The case of $\chi$ Cyg is also illustrative, the field ($B\textsubscript{E}$) being 0.7-4.3 Gauss, to be compared with the former estimate from \cite{Herpin2006} of 48.6 Gauss (0-85.4 Gauss, after re-evaluation of the field intensity). Considering the 2-3 Gauss measurement by \citet{Lebre2014} at the surface of $\chi$ Cyg, we could expect in the SiO layer a field strength of 0.5-1.5 or 0.1-0.8 Gauss if the field would vary as $1/r$ or $1/r^2$, respectively. Our results hence agree with a poloidal field, but do not rule out a toroidal component either, at least locally. This comparison has to be taken with caution, considering again that both measurements are well separated in time. The field strength variation observed in U Her over a 2-month period emphasizes this point. In addition, for the RSG $\mu$ Cep we derive upper magnetic field values larger that what \citet{Tessore2017} estimated at the stellar surface (1 Gauss). The fact that these two values are not compatible with a field strength decreasing with increasing distance may underline again the time variability of the magnetic field.

More generally, in Fig. \ref{fig:Bevol} we revisit all magnetic-field measurements gathered from the literature. Our new measurements provide new constraints for the SiO shell region at a few stellar radii and seem to exclude a global poloidal field, specially if we assume all maser emissions are saturated. Nevertheless, based on our $\chi$ Cyg observations, we cannot exclude a poloidal field close to the surface of this star evolving into a dominant toroidal field in the SiO maser region, as proposed by \citet{Meyer2021} in supergiant stars. 

The origin of this magnetic field is still debated, but mechanisms involving an $\alpha^2-\omega$ turbulent dynamo in AGB stars, and amplification by convection, stellar pulsation and cool spots can be invoked \citep[][]{Lebre2014}. As already proposed by \citet{Schwarzschild1975} for giant stars, the radial component of the field can then be strongly enhanced at a specific location near the surface above cool magnetic spots \citep[][]{Priest1987}. The tangential B field component can be amplified by shock compression due to the stellar pulsation \citep[][]{Hartquist1997}. Hence, a magnetic field  of 1-10 Gauss can be reached in the vicinity of cool magnetic spots, then decreasing in intensity to the $1/r^{2}$ law \citep[][]{Soker2002}. The same scenario applies in RSG stars with a stronger influence of their larger convective elements. Actually, as explained in Sect. \ref{sec:Icomp}, even the EVPA rotation seen in SiO features could be explained by a magnetic field orientation rapidly changing above cool spots.

The magnetic field variations observed in the AGB U Her over a two-months scale (stellar phases 0.25 and 0.45) suggest a possible link with the stellar phase, i.e. with pulsation$/$photospheric activity. It is interesting to note that for U Her, using ALMA, \citet{Baudry2023} have also identified and mapped strong time variations of the 268 GHz v2=2 line of water at phases 0.13, 0.80 and 0.92; this line is excited in regions of the inner CSE where the SiO masers are observed. We also mention that \citep[][]{Rosales2023} have seen with GRAVITY$/$VLTI structural changes at 3 $R_{\star}$ in the CSE of the Mira R Car (period= 310 days) on a one-month interval, probably linked to the photospheric activity. Quasi-simultaneous observations of the stellar photosphere (to detect cool spots) and polarimetry (to measure the field strength) are needed to prove any potential link. Actually, polarimetric interferometric observations with ALMA or the next-generation VLA would ideally allow us to spatially associate stellar spots with B-field enhanced regions. The role of pulsation could also be studied with a monitoring of the magnetic field with the stellar phase.

\section{Conclusions}
\label{sec:Conclusions}

We have performed new polarimetric SiO maser observations with the XPOL instrument at the IRAM-30m in a new sample of cool evolved stars. Applying a new data reduction methodology, we have removed most of the instrumental polarization to derive circular and linear polarization fractions with a high accuracy. From those and maser theory results, we obtain magnetic field strengths (ranging from a few Gauss up to several tens of Gauss) that better constrain the SiO shell region above the stellar photosphere. A global poloidal field can be excluded. 

Moreover, our observations of the magnetic field strength in U Her over a 2-month period suggest a possible link with the stellar phase, i.e. with pulsation$/$photospheric activity. 

Despite the advances accomplished here, the precise origin of the observed magnetic field at the surface and in the inner circumstellar layers of AGB stars and Red Supergiants still remains to be fully characterized. In particular, a monitoring of the magnetic field  to check the stellar-phase dependence of the polarization and quasi-simultaneous observations with VLTI and$/$or CHARA to detect cool spots and to follow the photospheric activity is under consideration. In addition, since the present results still suffer from moderate spatial resolution, ALMA polarimetric observations above the photosphere would also be most desirable.

\begin{acknowledgements}
This work is funded by the French National Research Agency (ANR) project PEPPER (ANR-20-CE31- 0002). MM acknowledges funding from the Programme Paris R\'egion fellowship supported by the R\'egion Ile-de-France. This project has received funding from the European Union’s Horizon 2020 research and innovation program under the Marie Skłodowska-Curie Grant agreement No. 945298. We would like to thank IRAM staff at Granada (Spain) for their support. We also thank the referee for his careful reading and his useful comments and suggestions. 
\end{acknowledgements}

\bibliographystyle{aa}
\bibliography{main.bib}

\begin{appendix}

\section{Removal of instrumental polarization}
\label{app:removal}

\begin{figure}
  \includegraphics[width=\columnwidth]{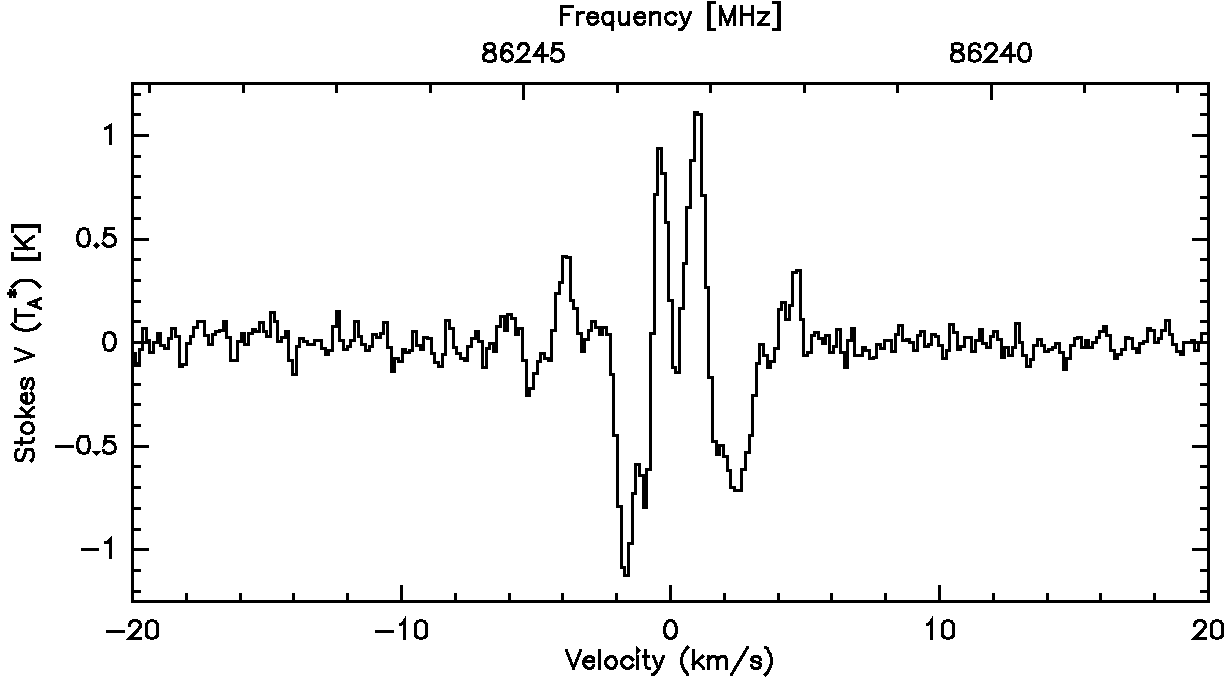}
  \includegraphics[width=\columnwidth]{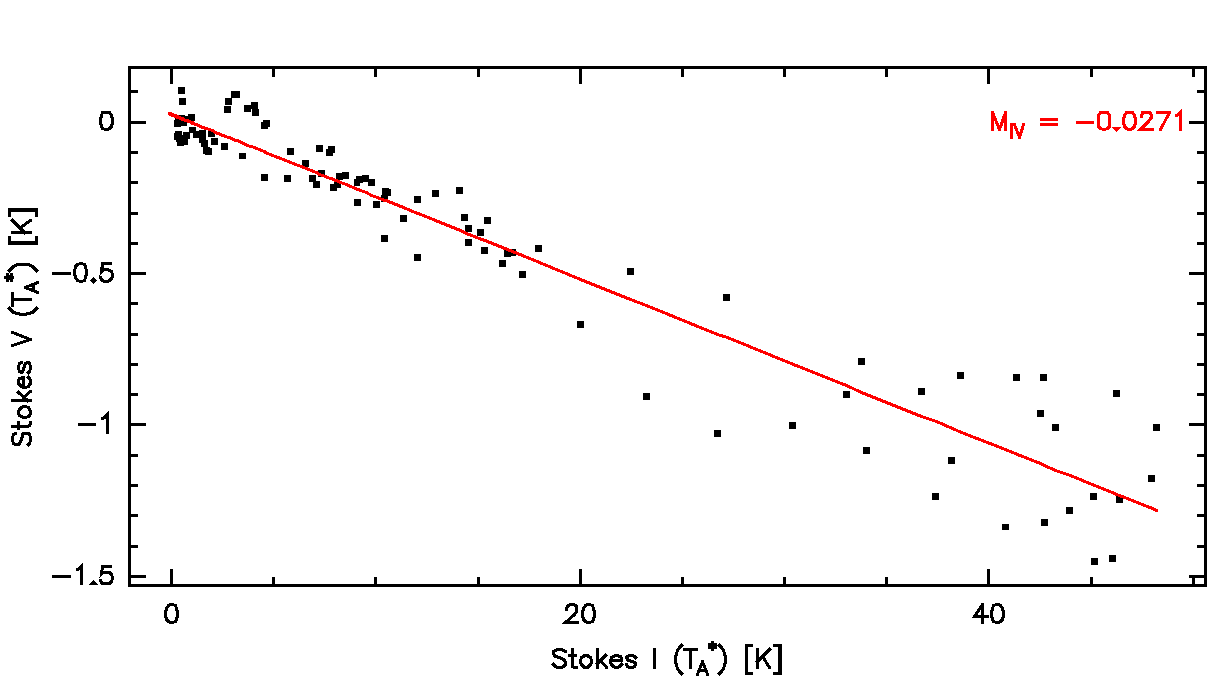}  
  \includegraphics[width=\columnwidth]{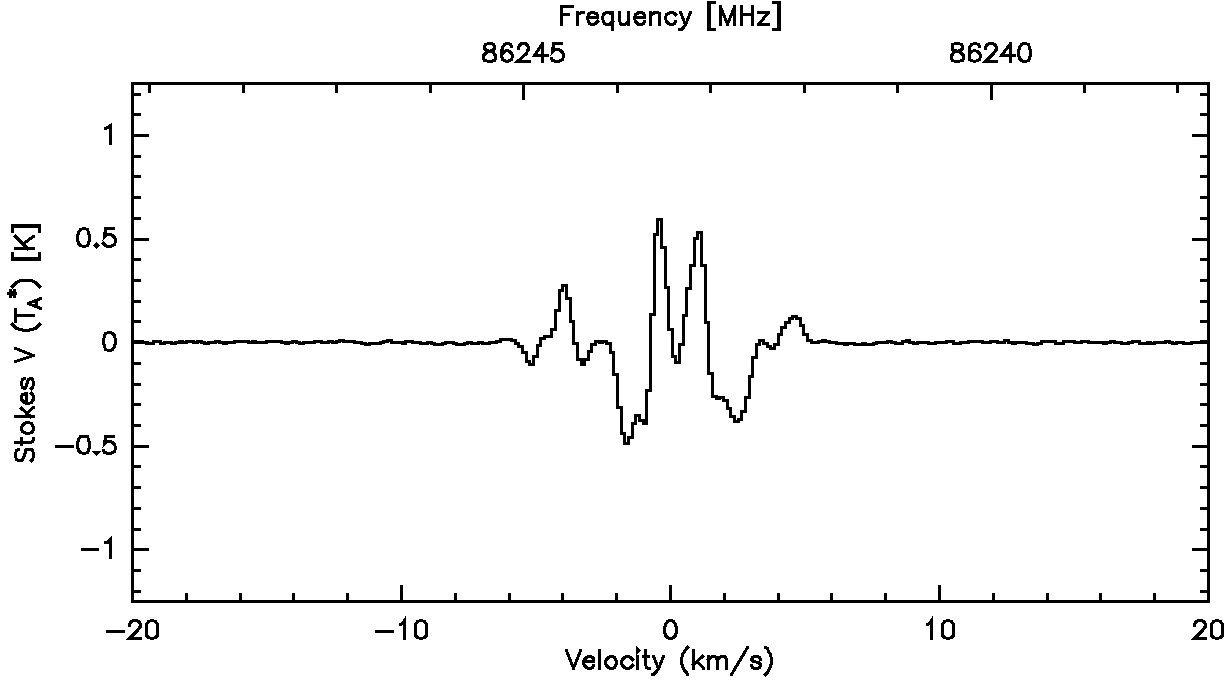}
    \caption{{\bf Top:} Average of Stokes $V$ spectra with low leakage ($|M_\mathrm{IV}| < 0.01$). {\bf Center:} Example of the correlation between $V_\mathrm{instr} = M_\mathrm{IV}\cdot I$ and Stokes $I$. The regression slope is $M_\mathrm{IV} =-0.0271$. {\bf Bottom:} Final cleaned and averaged Stokes $V$ spectrum. The vertical plot scale is the same as for the spectrum shown above.}
    \label{fig:Appen_ipv}
\end{figure}

\begin{figure}
    \includegraphics[width=\columnwidth]{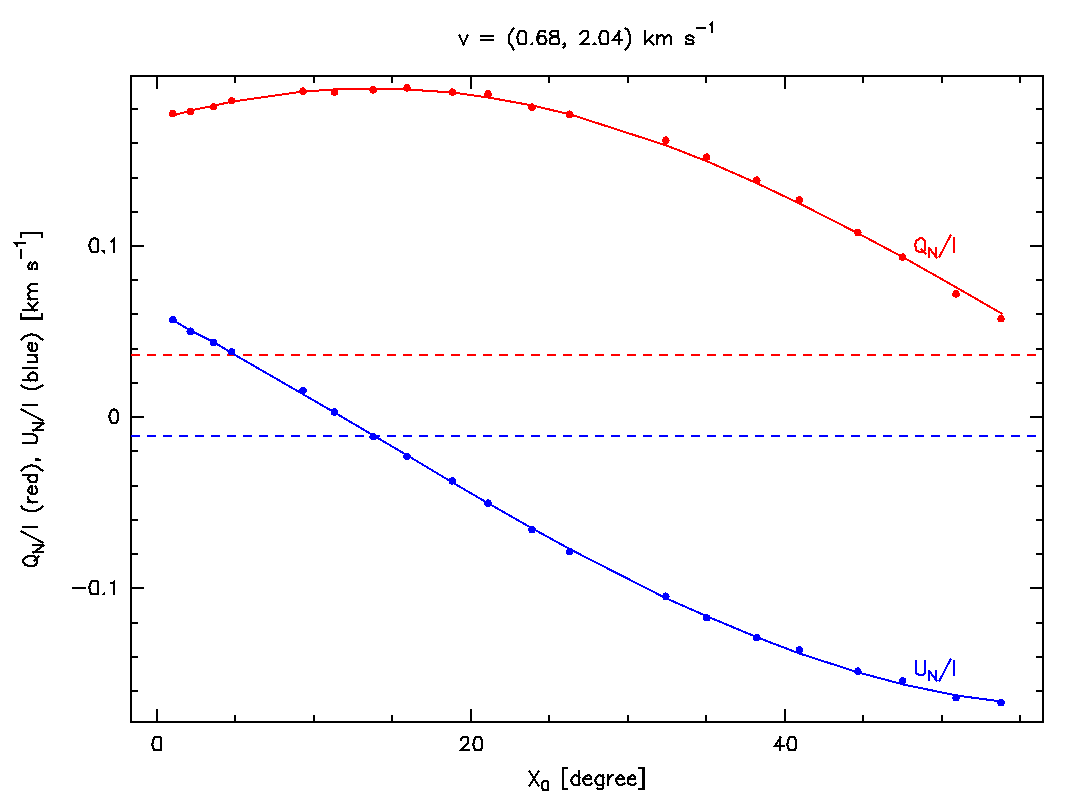}
      \caption{The solid red and blue lines show least-square fits to the fractional Stokes parameters $Q_\mathrm{N}/I$ and $U_\mathrm{N}/I$ (red, respectively blue, dots) of R~Leo versus $\chi_0 = \varepsilon-\eta$, averaged across a velocity interval framing the Stokes $I$ emission peak (0.68 to 2.04\,\kms) and expressed in the Nasmyth reference frame. The dashed lines indicate the corresponding fractional instrumental contributions. For details, see text.}
    \label{fig:Appen_ipqu}
\end{figure}

Removal of the instrumental leakage from Stokes $I$ to Stokes $V$, as described in Sect. \ref{sec:instr_I_V}, is demonstrated for R Leo in the three panels of Fig. \ref{fig:Appen_ipv}. The final cleaned $V$ spectrum (bottom panel) can be compared to the same $V$ spectrum  before cleaning (see Fig. \ref{fig:Stokes}). 
In Fig. \ref{fig:Appen_ipqu}, we demonstrate, again for the example of R Leo, the separation of the astronomical linear polarization from instrumental contributions, by fitting sine-waves, offset by the instrumental polarizations, to the fractional Stokes parameters $Q_\mathrm{N}/I$ and $U_\mathrm{N}/I$ (as defined in the Nasmyth system, i.e., with respect to the receiver cabin, see Sect. \ref{sec:instrUQ}), plotted against the angle $\chi_0$. 

\section{Stokes $I$ spectra}

\begin{figure*}
        \centering
        \includegraphics[width=1.9\columnwidth]{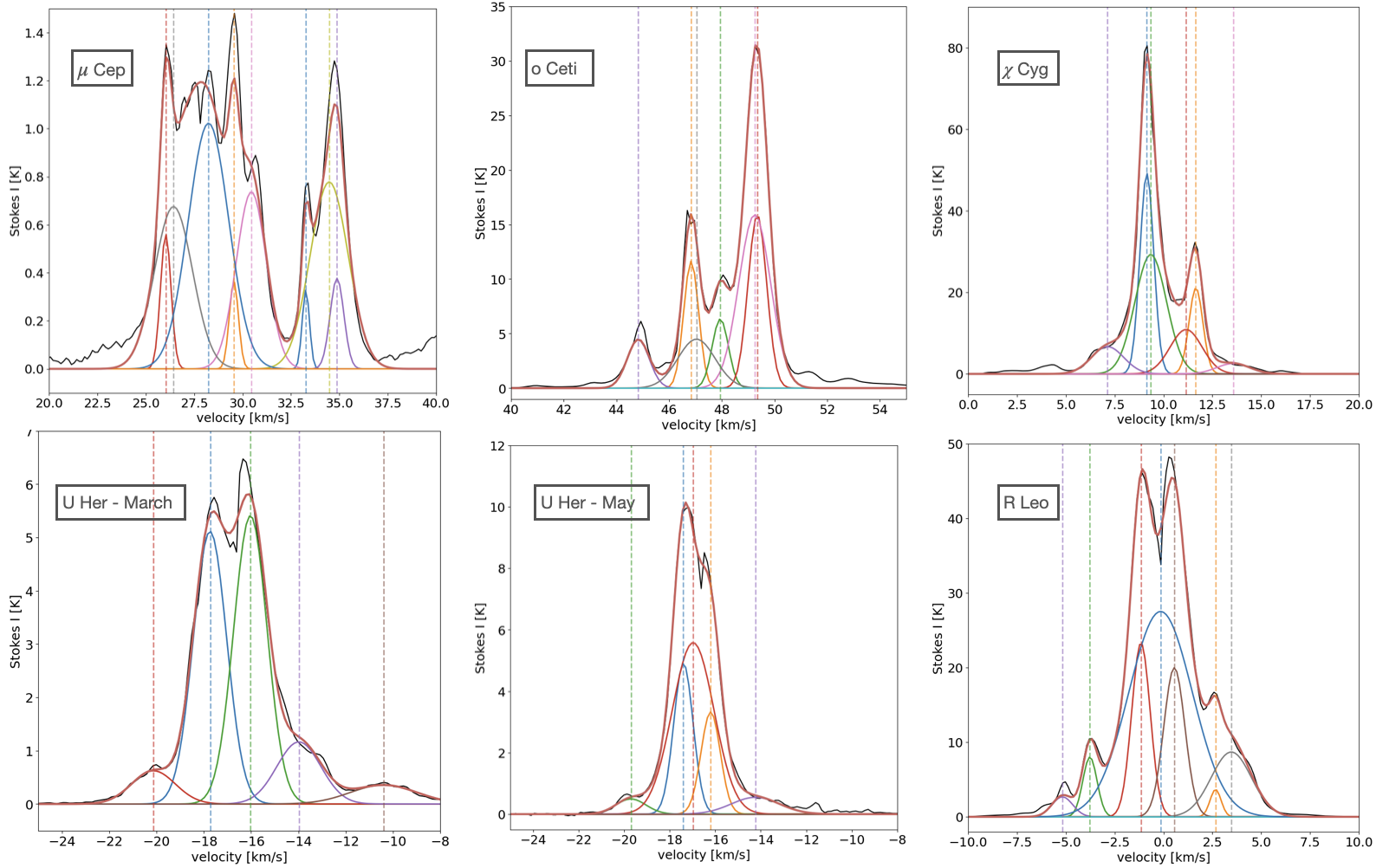}
        \caption{Decomposition of Stokes I for $\mu$ Cep, $o$ Ceti, $\chi$ Cyg, U Her, and R Leo (as for R Aql, see Fig.\ref{fig:raqldecompo}).}
        \label{fig:stokesI}
    \end{figure*}

Fig. \ref{fig:stokesI} presents the decomposition of Stokes $I$ for $\mu$ Cep, $o$ Ceti, $\chi$ Cyg, U Her (March and May 2022), and R Leo. The decomposition in R Aql is presented in Fig. \ref{fig:raqldecompo} (left panel).

\section{On the Land\'e factors of the SiO maser lines.}
\label{app:lande}

In this Appendix we recall some basic notions used to describe the Zeeman effect in molecules and define the appropriate value of the Land\'e factor, $g$, to be used in the SiO transition observed in this work.

The Larmor frequency can be used to describe the Zeeman shift of a level (in the energy domain) or a Zeeman transition (in the frequency domain) up to a gyromagnetic factor called the Land\'e factor. This factor is not explicitly dependent on the angular momenta that can be of two types. Those involving the electron produce an effect proportional to the Bohr magneton which, in the SI units, is given by
$$\mu_B= \frac{eh}{4\pi m_e},$$ 
and those involving the whole molecule for which the nuclear magneton is relevant
$$\mu_N= \frac{eh}{4\pi m_p}.$$ 
Both magnetons are related by the proton-to-electron mass ratio $m_p/m_e=1836$.

The molecule under study here, SiO, is diamagnetic (its ground electronic state is $^1\Sigma ^+$) and, consequently, all its electronic angular momenta are zero. It is the Larmor frequency in terms of the nuclear magneton that must be used:
$$\nu_L=\frac{\mu_N}{h}B=762.31 \cdot B[G]$$
(for reference, we recall the common Larmor frequency in terms of the Bohr magneton $\nu_L=1.3996\times 10^{6}\cdot  B[G]).$ \\

Calling Zeeman shift, the shift of an energy level with respect to its zero-field position, one defines the Zeeman splitting as the separation between two, usually symmetric, Zeeman split components using the selection rule $\Delta M_J=\pm 1$ or 0  where $M_J$ is the projection of momentum J along the magnetic field axis (excluding 0 to 0 for $\Delta J=0$). Laboratory measurements \citep{Davis1974, tischer74} are coincident in assigning $g\approx 0.155$ for all J rotational transitions of SiO in the first vibrational state v=1. The Land\'e factor is slightly different, 0.154 in the v = 0 state (not observed here). (There is also a weak magnetic field dependent energy term due to the magnetic susceptibility anisotropy of the SiO molecule \citep{Davis1974}; it can be numerically ignored here.) The Zeeman shift of the v=1 SiO transitions, in terms of the nuclear magneton, is thus $g\nu_L=118.16 \cdot B[G]$ and the Zeeman splitting is twice that value. 

In the present work, we have defined an explicit Land\'e factor including the appropriate correction from the Bohr magneton to the nuclear magneton as 
$$g_B = \frac{1}{1836}g = 8.4\cdot 10^{-5}.$$
We note that this was also used in \citet{Elitzur1996} (see his dimensionless equation 1.1 where, however, the Land\'e factor must account for the 1/1836 factor for diamagnetic molecules -Elitzur’s Doppler width is also half the total Doppler line width at half intensity), and also used in \citet{Kemball_1997, Kemball2009} or in \citet{Asensio2005}.

Finally, we point out that starting from \citet{Goldreich1973}, the concept of the Zeeman frequency,  $g\Omega$, was introduced in many subsequent works on the magnetic field intensity where $\Omega$ is in fact the proton gyro frequency. In this work the Zeeman frequency is just the Zeeman splitting and, hence, twice the Zeeman shift $$g\Omega = 2g\nu_L.$$
For the SiO lines of interest, using now $\nu_L$ in terms of the nuclear magneton we have in Hz
$$ g\Omega = 1524.6gB[G]$$ (and $2799165g_B B[G]$ Hz in terms of the Bohr magneton).

\section{Magnetic Field Strength}
\label{sec:Appen:MF}

\begin{figure*}
\centering
        \includegraphics[width=0.6\columnwidth]{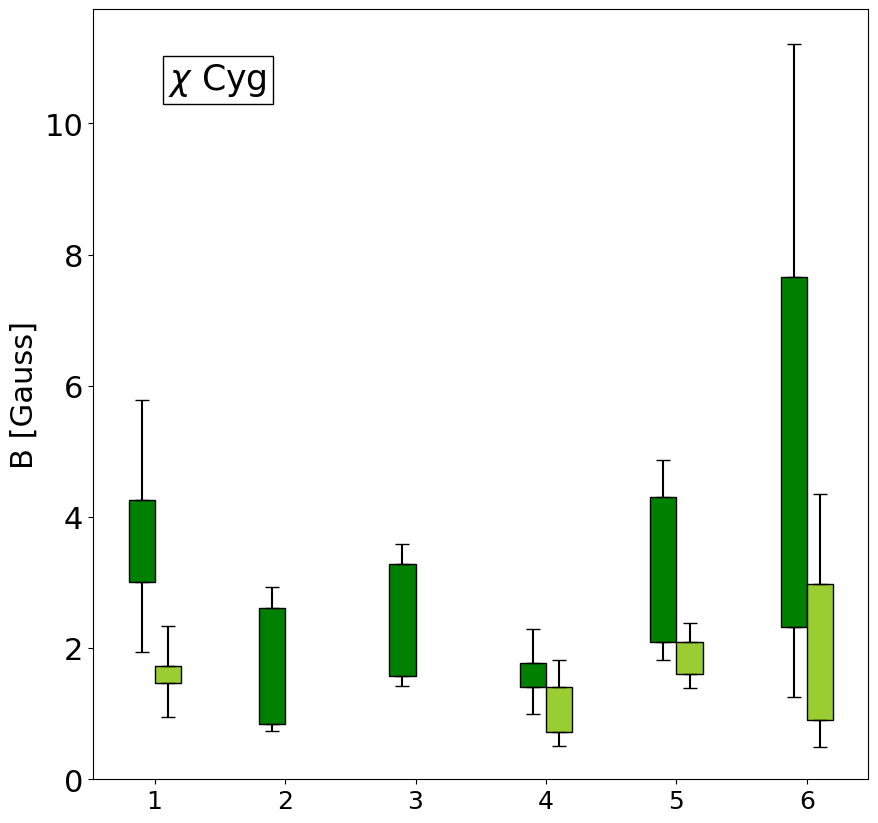}
        
        \caption{Magnetic field values for $\chi$ Cyg. Same as Fig. \ref{fig:Uher}.}
         \label{fig:Appen_sat}
\end{figure*}

\begin{figure*}
\centering
        \includegraphics[width=0.6\columnwidth]{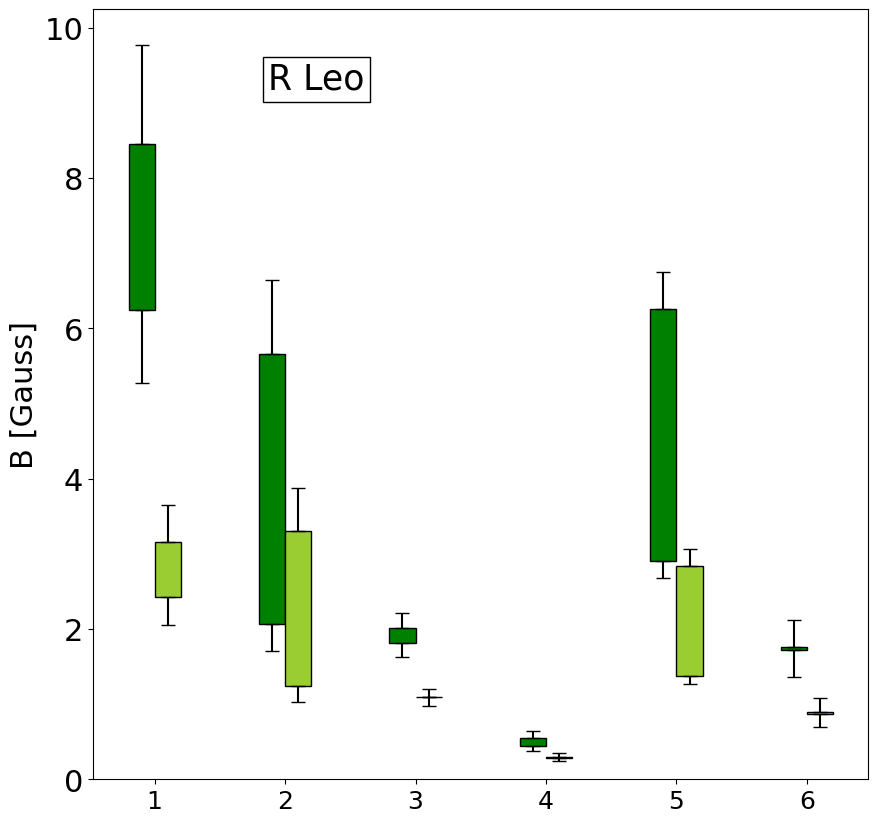}
        \includegraphics[width=0.6\columnwidth]{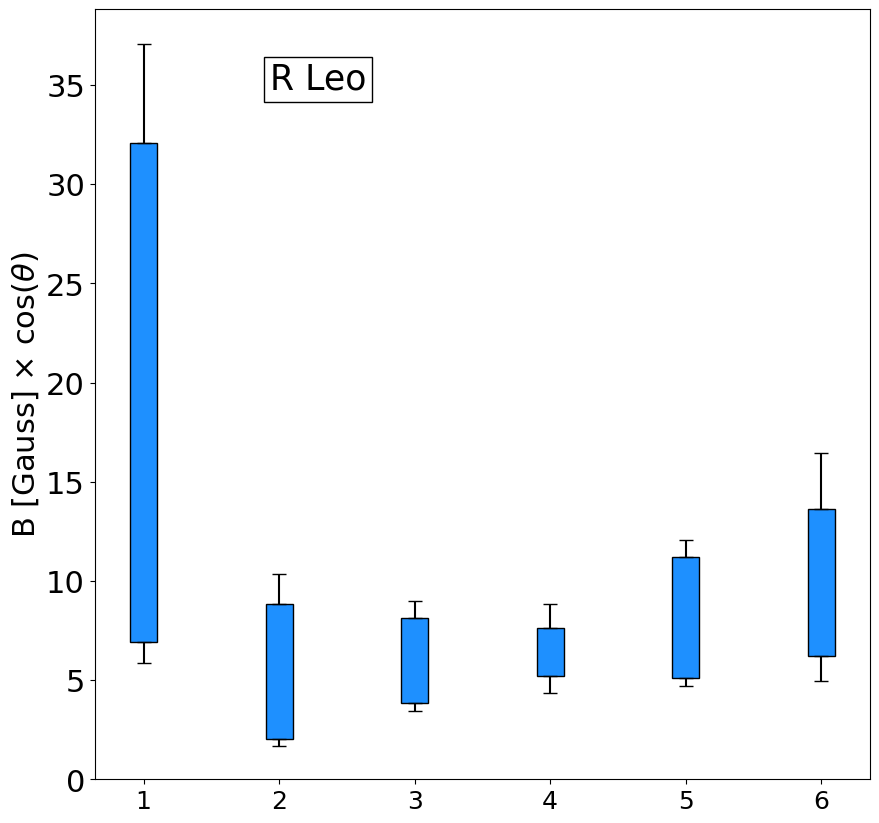}\\
        \includegraphics[width=0.6\columnwidth]{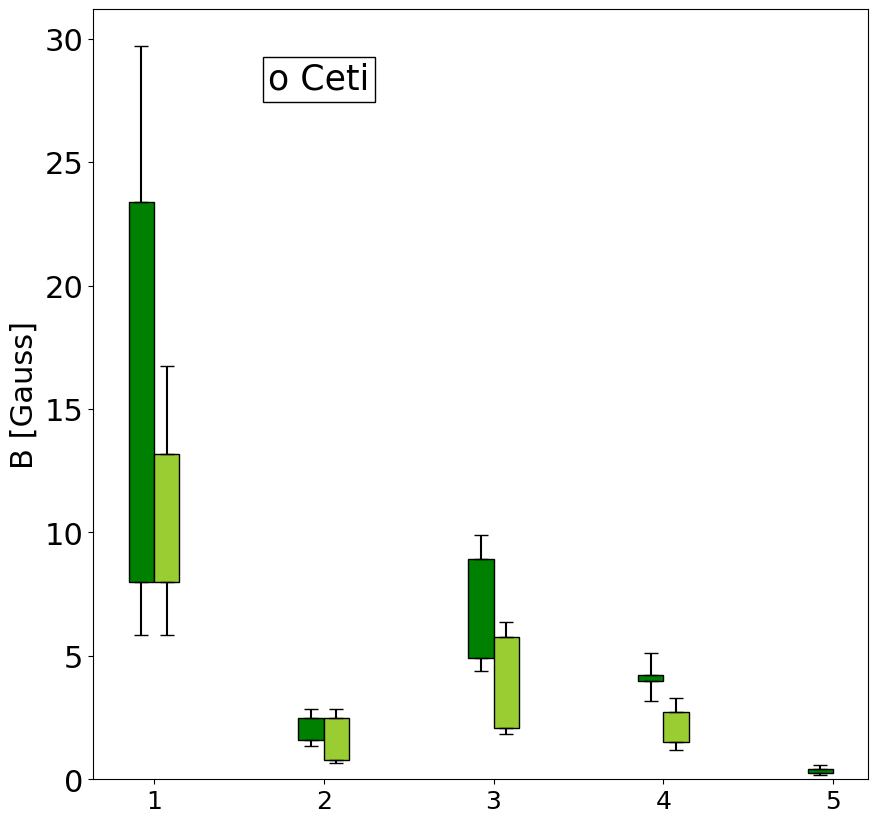}
        \includegraphics[width=0.6\columnwidth]{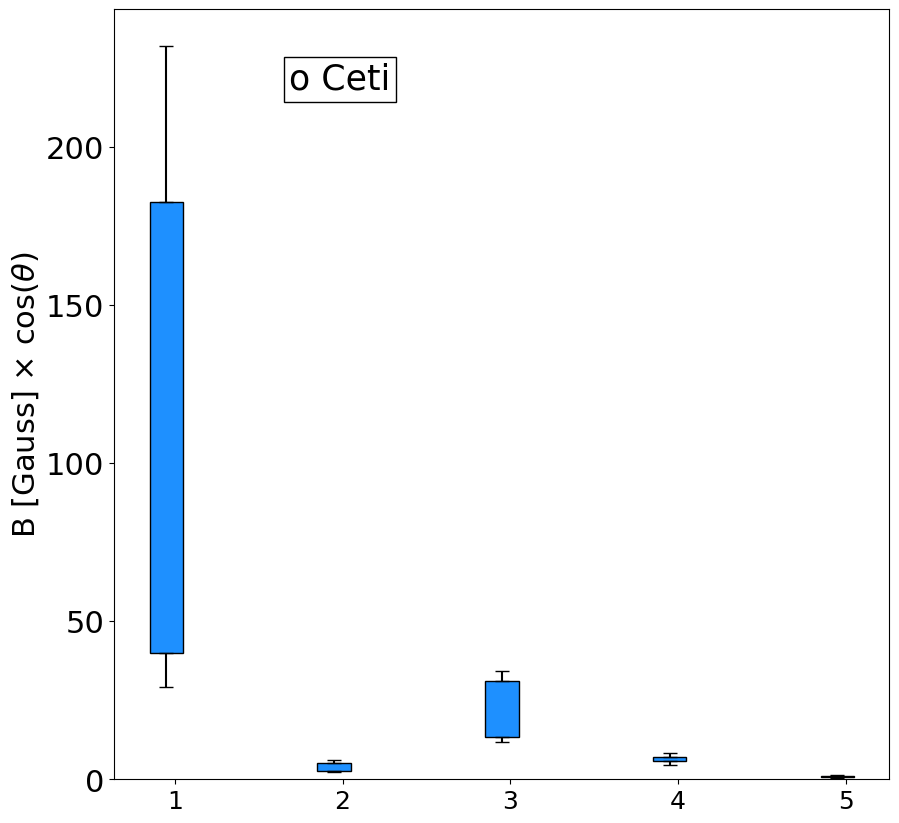}
\caption{Magnetic field values for sources with uncertain saturation status. In green, the magnetic field computed with the Elitzur's formula. In blue, the magnetic field from the Zeeman weak-field theory ($B\times \textrm{cos}(\theta_2)$).}
         \label{fig:Appen_unc}

\end{figure*}

Fig. \ref{fig:Appen_sat} presents the magnetic field strengths of $\chi$ Cyg, using Eq. \ref{eq:Elitzur} for saturated masers with two solutions if the linear polarization is under $33\%$. Figure \ref{fig:Appen_unc} presents the magnetic field strengths for R Leo and $o$ Ceti for saturated masers from Eq. \ref{eq:Elitzur} (left panel) and for unsaturated maser from Eq. \ref{eq:Zeeman} (right panel). Table \ref{tab:pmoyen} shows the average circular polarization fraction $\overline{p_\mathrm{C}}$ and average ${\overline{p_\mathrm{L}}}^2/4$ for each star.

\begin{table}
\centering
\caption{Average circular polarization fraction $\overline{p_\mathrm{C}}$ and average ${\overline{p_\mathrm{L}}}^2/4$ for each star.} 
\label{tab:pmoyen}
\begin{tabular}{ccc}
\hline
\hline
    Star & $\overline{p_\mathrm{C}}$ & ${\overline{p_\mathrm{L}}}^2/4$ \\
    \hline
R Aql & 0.022 & 0.005\\
$\mu$ Cep & 0.042 & 0.001\\
$o$ Ceti & 0.022 & 0.019\\
$\chi$ Cyg & 0.013 & 0.028 \\
U Her & & \\
(March)& 0.009 & 0.002\\
(May) & 0.012 & 0.004\\
R Leo & 0.013 & 0.006\\
\hline
\end{tabular}
\end{table}

\end{appendix}

\end{document}